\documentclass[letterpaper,9pt]{article}
\usepackage{tabularx} 
\usepackage{amsmath}  
\usepackage{graphicx} 
\usepackage[margin=1in,letterpaper]{geometry} 
\usepackage[final]{hyperref} 
\usepackage[url=false, doi = false, sorting=none]{biblatex}
\usepackage{dirtytalk}
\usepackage{amssymb}
\usepackage{authblk}
\usepackage{url}
\usepackage{xcolor}

\hypersetup{
	colorlinks=true,       
	linkcolor=blue,        
	citecolor=blue,        
	filecolor=magenta,     
	urlcolor=blue         
}
\newcommand{\bb}[1]{\boldsymbol{#1}}

\begin{document}

\title{An Auto-Differentiable Likelihood Pipeline for the Cross-Correlation of CMB and Large-Scale Structure due to the Kinetic Sunyaev-Zeldovich Effect}

\author{Yurii Kvasiuk{\href{https://orcid.org/0009-0002-4720-1320}{\includegraphics[height=8pt]{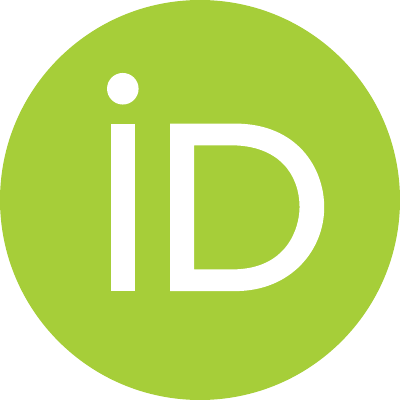}}}\thanks{Corresponding author. Email: kvasiuk@wisc.edu} }
\author{Moritz M\"unchmeyer{\href{https://orcid.org/0000-0002-3777-7791}{\includegraphics[height=8pt]{ORCIDiD_iconvector.pdf}}}}
    
\affil{Department of Physics, University of Wisconsin-Madison, Madison, WI 53706, USA}

\date{\today}

\maketitle
\begin{abstract}
We develop an optimization-based maximum likelihood approach to analyze the cross-correlation of the Cosmic Microwave Background (CMB) and large-scale structure induced by the kinetic Sunyaev-Zeldovich (kSZ) effect. Our main goal is to reconstruct the radial velocity field of the universe. While the existing quadratic estimator (QE) is statistically optimal for current and near-term experiments, the likelihood can extract more signal-to-noise in the future. Our likelihood formulation has further advantages over the QE, such as the possibility of jointly fitting cosmological and astrophysical parameters and the possibility of unifying several different kSZ analyses. We implement an auto-differentiable likelihood pipeline in JAX, which is computationally tractable for a realistic survey size and resolution, and evaluate it on the Agora simulation. We also implement a machine learning-based estimate of the electron density given an observed galaxy distribution, which can increase the signal-to-noise for both the QE and the likelihood method. 
\end{abstract}

\section{Introduction}
The kinetic Sunyaev-Zeldovich (kSZ) effect is the dominant contribution to the total blackbody Cosmic Microwave Background (CMB) anisotropies at small scales, overcoming CMB lensing around $\ell \simeq 4000$. It was predicted in the early 70s \cite{kSZ_prediction} and first detected with the Atacama Cosmology Telescope (ACT) in 2012 \cite{Hand_2012}. The kSZ effect is a Doppler shifting of CMB photons due to the Thompson scattering on free electrons on the line of sight. The kSZ signal is both sensitive to bulk velocities as well as to galaxy and cluster astrophysics. Measurements of the kSZ are able to constrain different dark energy models \cite{dedeo_spergel_2005, Hernandez_Monteagudo_2006,Bhattacharya_2008}, neutrino masses \cite{2015PhRvD..92f3501M, 2020JCAP...07..043K}, modified gravity \cite{Mak_2012, Pan_2019} and compensated isocurvature perturbations (CIP) \cite{arxiv.2208.02829}. Further, the kSZ signal can be used to tightly constrain inflationary models that predict so-called local primordial non-Gaussianity $f_{NL}$ \cite{M_nchmeyer_2019, arxiv.2205.03423}. 

To use the kSZ signal for cosmological constraints, it needs to be combined with a tracer of the electron density field, for example using a galaxy survey. By cross-correlating the temperature measurement from a CMB experiment with the galaxy field from a large-scale structure survey, one can reconstruct the radial velocity (or momentum) field $v_r$ on large scales \cite{Deutsch_2018}. One can then convert this measurement to a measurement of the radial matter over-density using the continuity equation $\delta_m(k) \sim k^2v_r(k)$, and obtain an extremely low noise measurement of this field \cite{arxiv.1810.13423}.

The canonical way to estimate $v_r$ from the kSZ effect is to construct an analytic quadratic estimator (QE) in spherical harmonics or Fourier space \cite{arxiv.1810.13423, Giri_2022, Deutsch_2018, Cayuso_2021}. Studies of this method on simulations were presented in \cite{Cayuso:2018lhv,Giri:2020pkk}. Prior to the QE approach, other kSZ estimators that are sensitive to the velocity field were developed also in \cite{Terrana:2016xvc,2010MNRAS.407L..36Z,Zhang:2015uta}. In the present work, we develop an alternative maximum likelihood approach to kSZ velocity reconstruction. The possible estimators for kSZ velocity reconstruction are similar to those of gravitational lensing in the CMB. In fact, the QE was first introduced in \cite{Hu:2001kj} for CMB lensing. In CMB lensing, the QE is optimal at low signal-to-noise, but at higher signal-to-noise (in particular where the total CMB anisotropy is dominated by lensing) it is substantially suboptimal. For this reason, several authors have developed a maximum likelihood approach to CMB lensing, which is optimal at all noise levels, but comes at the cost of a computationally expensive optimization process. The first maximum likelihood estimator for lensing was developed in \cite{Hirata:2002jy}. A closely related method and code implementation (\texttt{LensIt}) to find the maximum a posteriori (MAP) iteratively was presented in \cite{Carron:2017mqf}. A different lensing likelihood was presented in \cite{Millea:2017fyd,Millea:2020cpw}, with associated code \texttt{CMBLensing.jl}, and recently applied to SPT data using HMC to sample the likelihood \cite{Millea:2020iuw}. For the small high-resolution sky patch analzyed in \cite{Millea:2020iuw}, it was shown that the likelihood outperforms the QE on real data. For CMB S4, it is expected that lensing likelihood methods can improve the signal-to-noise over the QE by up to a factor of two \cite{Legrand:2021qdu}.

The goal of our present work is to develop a similar likelihood approach to kSZ velocity reconstruction, or more generally, to the cross-correlation of the kSZ signal with a large-scale structure survey. As we will see, at high enough signal-to-noise, the likelihood again outperforms the quadratic estimator substantially. There are however some important differences compared to the CMB lensing likelihood. CMB lensing is a non-linear re-mapping of the primary CMB temperature, while the kSZ, at leading order, is linearly added to the primary CMB. This should make it easier to optimize the likelihood. Further, one needs to model the relation of the galaxy distribution and the electron distribution in the likelihood, which can only be approximated. Recently a different maximum likelihood kSZ estimator was proposed in \cite{arxiv.2205.15779}. However, this estimator is not optimization-based and instead makes some analytic assumptions that allow for an analytic solution. We will comment below in more detail on the relation to this prior work. Another related work is \cite{Nguyen:2020yuc}, which developed a pipeline to detect the kSZ signal using a forward modeling-based velocity template and a kSZ model at the cluster level. 

A number of autodifferentiable forward models for cosmology have recently been proposed in other contexts. In \cite{Bianchini:2022wte} an autodifferentiable likelihood is used to analyze gravitational lensing of the CMB and patchy screening during the epoch of reionization. For the study of LSS formation, pmwd \cite{Li:2022qlf}, a differentiable cosmological particle mesh, implemented with JAX, has been presented. In the context of weak gravitational galaxy lensing, recent developments include MADLens \cite{Bohm:2020ilt}, a package for autodifferentiable calculation of non-Gaussian convergence maps, and Differentiable Lensing Lightcone (DLL) \cite{Lanzieri:2023ftk}, a GPU-based differentiable weak lensing forward model which is differentiable with respect to all cosmological parameters.

In the present work we develop an autodifferentiable approach to kSZ analysis and aim to answer the following main questions:
\begin{itemize}
    \item What likelihood is the most useful for our purpose, and what choices can be made? We perform a separation of scales between the large-scale (linear) velocity field and the small-scale galaxy/electron distribution, which avoids the need for non-linear forward modeling of large-scale structure. We also discuss different assumptions for the statistical relation of galaxies and electrons.
    \item Can we consider a realistically sized data set (angular resolution and redshift bins) and find the maximum a posteriori (MAP) effectively on current hardware? To achieve this, we implement the likelihood using the auto-differentiable language JAX, which provides a highly optimized GPU implementation. 
    \item How does the signal-to-noise in the likelihood compare with the quadratic estimator as a function of experimental parameters? As we discuss below, for realistic current and next-generation data the kSZ likelihood can probably not outperform the QE (see below for details), however, we show that for more futuristic experiments the improvement can be substantial, despite the presence of high redshift kSZ from reionization. 
    \item Can machine learning be used to raise the signal-to-noise by learning an improved template of the electron distribution given observed galaxies, by training on simulations? While a full exploration of this idea requires a study on hydrodynamic simulations, we show encouraging initial results using the matter distribution, that could already improve the velocity reconstruction for Simons Observatory. 
\end{itemize}

We make forecasts for a number of idealized experimental configurations. Our simulation study is not fully realistic, since we do not take into account effects such as photo-z errors (which would degrade forecasts) or halo mass tracers (which could improve them). For the upcoming Simons Observatory (SO) or CMB-S4 combined with Rubin Observatory we do not predict a signal-to-noise improvement of the likelihood over the QE, but our results are not fully conclusive because we have not analyzed a simulation with the full halo density of Rubin Observatory. However, even in the case where the QE is statistically optimal, an optimization-based formulation has several advantages over the QE, some of which we will study in future work:
\begin{itemize}
    \item The likelihood can include both cosmological and astrophysical parameters, and it becomes thus possible to jointly fit both of these and obtain combined constraints with covariances. In the present work, we do not yet implement such a joint fit, but this is a main goal for future work. This could also include a model of the reionization kSZ.
    \item A likelihood approach can naturally take into account systematic uncertainties such as calibration issues, because it allows to fit a joint model to the data that takes into account all known experimental effects. This was recently demonstrated for CMB lensing in \cite{Millea:2020cpw}.
    \item The kSZ likelihood can include several different estimators. For example, the likelihood directly provides us with a partially kSZ-cleaned primary CMB map. So-called de-kszing has recently been studied in \cite{Foreman:2022ves} using a template method. The signal extracted by squared kSZ statistics (projected fields) \cite{Hill:2016dta}, which can probe the baryon distribution, is also included in the likelihood formulation because the likelihood includes a complete probabilistic model of all relevant fields. 
    \item It is conceptually straightforward (but practically and computationally difficult) to combine several CMB secondary effects into an overall CMB x LSS likelihood. A joint lensing-kSZ likelihood could estimate both the lensing potential and the radial velocities, include the ISW effect (e.g. \cite{Dong:2020fqt}) and the moving lens effect (e.g. \cite{Hotinli:2018yyc}), include lensing of the kSZ, and include non-blackbody contributions such as the thermal Sunyaev Zeldovich effect. 
    \item In the likelihood formulation we can use non-Gaussian priors that capture the non-Gaussian small-scale statistics of the galaxy and electron density. A non-Gaussian prior could be implemented for example with a normalizing flow \cite{Rouhiainen:2022cwd,Rouhiainen:2021qjg}, trained on hydrodynamic simulations, and can increase the signal-to-noise over a Gaussian prior, since these fields are substantially non-Gaussian.    
    \item The kSZ likelihood can be included in a more general forward modeling setup. In forward modeling of large-scale structure, one starts with the Gaussian initial conditions and then maps them to late-time observables with a differentiable forward model of structure formation (see e.g. \cite{Seljak:2017rmr,2013MNRAS.432..894J}). Radial velocities were recently used for reconstruction of the initial conditions in \cite{Bayer:2022vid}.     
    \item Once a likelihood is available, one can in principle not only find the MAP but also perform a fully Bayesian analysis, for example using variants of Hamiltonian Monte Carlo (HMC) or Variational Inference (VI). It seems unlikely that this is computationally tractable at the full resolution of the data, but it may be possible to develop useful approximations, for example by approximately integrating out the small-scale fields. 
\end{itemize}

The paper is organized as follows. In Sec. \ref{sec:qe} we set up our notation and review the conventional quadratic estimator formulation of kSZ velocity reconstruction in our coordinates. In Sec. \ref{sec:likeli} we discuss the kSZ likelihood, its relation to the QE, its implementation, and how to find the MAP of field parameters and scalar parameters. In Sec. \ref{sec:results} we apply our method to simulations and evaluate the improvement factor over the QE. In Sec. \ref{sec:machinelearning} we discuss a machine learning method, which can improve the velocity reconstruction for either the QE or the likelihood. Our conclusions and future work are summarized in Sec. \ref{sec:conclusion}.

\section{Review of the Quadratic Estimator approach}
\label{sec:qe}

We start by reviewing the standard kSZ velocity reconstruction. For computational simplicity, we will work in binned flatsky coordinates. This is appropriate in particular for a photometric survey such as Rubin Observatory. Redshift binning does not lose significant information compared to a full 3-dimensional analysis if the photometric error bars are larger than the redshift binning. 

\subsection{Notation}
We adopt the following Fourier conventions for scalar fields in a 2D flatsky Cartesian basis: 
\begin{align}
    f(\bb{x}) = \frac{1}{(2\pi)^2}\int e^{i\bb{l}\bb{x}}f(\bb{l})d^2l 
    \hspace{2cm} f(\bb{l}) = \int e^{-i\bb{l}\bb{x}}f(\bb{x})d^2x.
\end{align}
 The 2D power spectrum $P^{f}(l)$\footnote{Relations between 2D flatsky, 2D angular, and 3D power spectra are reviewed in Appendix \ref{appdx:a1} and \ref{appdx:a2}} is defined as 
 \begin{align}
 \langle f(\bb{l}_1)f^*(\bb{l}_2)\rangle = (2\pi)^2\delta^{(2)}(\bb{l}_1-\bb{l}_2)P^{f}(l_1)\end{align}
In this work, we will use $\frac{\Delta T}{T}\equiv \theta$ for the CMB anisotropy; matter (galaxy, electron) overdensity is denoted by $\delta^{m,(g,e)}\equiv \rho^{m,(g,e)}/\bar{\rho}^{m,(g,e)}-1$ and should not be confused with $2D$ Dirac delta-function $\delta^{2D}(\bb{x})$ or Kronecker symbol $\delta^{(K)}_{\alpha,\beta}$. The radial velocity is denoted by $v_r$ and sometimes we will omit the superscript $r$ for notational simplicity.
The data required in this work is the CMB temperature fluctuation $\theta(\bb{l})$ and the binned galaxy field $\delta^\alpha_g(\bb{l})$. 

\subsection{Binned kSZ signal}
\label{binned_ksz}

The temperature anisotropy generated by the kSZ is proportional to the electron density and the velocity and is given by the following line-of-sight integral:
\begin{equation}
\theta(\hat{n}) \big|_{kSZ}  = - \sigma_T \int d\chi \ a \ n_e(\hat{n},\chi) v_r(\hat{n},\chi),
\end{equation}
over comoving distance $\chi$. In this equation, $\sigma_T$ is the Thomson scattering cross section, $a$ is the scale factor ($a(z=0)=1$) and $n_e$ is the electron number density. We now assume that we can probe the electron density $n_e$ with a finite radial resolution, given roughly by the redshift error of the experiment (or below by the radial binning of the available simulations). Given a set of radial bins indexed by $\alpha$ (60 bins for $0.5<z<3$ in our main analysis), we can write the binned kSZ as \cite{Deutsch_2018}
\begin{equation}
\label{eq:ksz_binned}
    \theta(\bb{x}) \big|_{kSZ}^{binned} = \sum_{\alpha=1}^{N_{bins}}\tau^{\alpha}(\bb{x})v_r^{\alpha}(\bb{x})
\end{equation}
Here $\tau^{\alpha}(\bb{x})$ and $v^{\alpha}(\bb{x})$ are the integrated optical depth and radial velocity in the bin $\alpha$ as functions of 2D Cartesian coordinate $\bb{x}$. We further assume that 
\begin{equation}
\label{eq:tau}
\tau^{\alpha}(\bb{x}) = f^{\alpha}(1+\delta^{\alpha}_e(\bb{x}))
\end{equation}
where $\delta_e$ is the electron overdensity integrated over the bin. Our binning is defined with a top-hat window function with the width of the bin $\Delta_{\alpha}$: 
\begin{equation}
    \delta^{\alpha}_m = \frac{1}{\Delta_{\alpha}}\int_{z^*_\alpha-\Delta_{\alpha}/2}^{z^*_\alpha+\Delta_{\alpha}/2} dz \ \delta_m(z)
\end{equation}
The binned kSZ from the radially averaged $\tau^{\alpha}$ and $v_r^{\alpha}$ do not constitute the total kSZ that an experiment will see. Small-scale fluctuations within the bin will contribute additional kSZ, which however cannot be described by the binned data. To have a realistic kSZ power in the map, including contributions from smaller radial scales than the bin width, occasionally we will add an additional \say{uncorrelated} kSZ component. This component also includes kSZ from redshifts where we have no galaxy data, including reionization. Thus we model the total kSZ temperature anisotropy as
\begin{align}
    \theta(\bb{x}) \big|_{kSZ}^{total} = \theta(\bb{x})^{kSZ, binned} + \theta(\bb{x})^{kSZ, uncorr}
\end{align}
We model $\theta(\bb{x}) \big|_{kSZ}^{uncorr}$ as a Gaussian random field which is uncorrelated with the binned galaxy and velocity fields, and adjust its power spectrum to have a realistic total kSZ amplitude in our maps. The total measured CMB temperature is thus
\begin{equation}
\label{eq:totalcmb}
\theta_{obs}(\bb{l}) = \theta(\bb{l})^{pLCMB} + \theta(\bb{l})^{kSZ, binned} + \theta(\bb{l})^{kSZ, uncorr} + n_\theta(\bb{l})
\end{equation}
which includes contributions from lensed primary CMB, kSZ effect, and noise. We write $\tilde{P}^{\theta}(l)$ for its corresponding power spectrum.

\subsection{Flat Sky Quadratic Estimator for the velocity field}

We now derive the quadratic estimator in binned flatsky coordinates (see App. \ref{appdx:a}), which closely follows the binned spherical harmonics derivation in \cite{Deutsch_2018}. The kSZ-induced correlation between total CMB temperature perturbation field $\theta_{obs}(\bb{l})$ 
and galaxy overdensity bin $\delta^{g \alpha}_{obs}(\bb{l'})$ (the tildes here mean including instrumental noise) is 

\begin{equation}
\begin{split}
    \langle \theta_{obs}(\bb{l_1})\delta_{obs}^{g,\alpha}(\bb{l_2})\rangle & = \sum_{\beta}\frac{1}{(2\pi)^4}\int d^2l'd^2l'' (2\pi)^2 \delta^{2D}(\bb{l_1-l'-l''})v_{\beta}(\bb{l''})\langle \tau_{\beta}(\bb{l'}) \delta^{g,\alpha}(\bb{l_2}) \rangle \\
    & = \int d^2l'd^2l'' \delta^{2D}(\bb{l_1-l'-l''})v_{\beta}(\bb{l''})P^{\tau g}_{\alpha}(l') \delta^{2D}(\bb{l_2+l'}) \\
    & = \int d^2l \delta^{2D}(\bb{l_1+l_2-l})v_{\alpha}(\bb{l})P^{\tau g}_{\alpha}(l_2)
\end{split}
\end{equation}
Where, under the assumption of independent bins, we defined 
\begin{align}
\langle \tau_{\beta}(\bb{l'}) \delta^{g}_{\alpha} (\bb{l}) \rangle
= (2\pi)^2P^{\tau g}_{\alpha}(l)\delta^{(K)}_{\alpha\beta}\delta^{2D}(\bb{l+l'})
\end{align}
The general quadratic estimator is then of the form
\begin{equation}
\hat{v}_\alpha(\bb{l}) = \int d^2l_1d^2l_2W_{\alpha}(\bb{l_1},\bb{l_2})\theta_{obs}(\bb{l_1})\delta_{obs}^{g,\alpha}(\bb{l_2})\delta^{2D}(\bb{l-l_1-l_2})
\end{equation}
We demand that $\langle\hat{v}_\alpha(\bb{l})\rangle = v_\alpha(\bb{l})$ and look for $\hat{v}_\alpha(\bb{l})$ that has smallest possible variance under the assumption that fields are Gaussian. That yields:
\begin{equation}
    W_{\alpha}(\bb{l_1},\bb{l_2}) = \lambda(l) \frac{P_{\alpha}^{g\tau}(l_2)}{\tilde{P}^{\theta}(l_1)\tilde{P}^{g}_{\alpha}(l_2)}
\end{equation}
where $\lambda(l)$ is a normalization:
\begin{equation}
    \lambda^{-1}(l) = \int d^2l_1d^2l_2 \delta^{2D}(\bb{l-l_1-l_2})\frac{(P_{\alpha}^{g\tau}(l_2))^2}{\tilde{P}^{\theta}(l_1)\tilde{P}^{g}_{\alpha}(l_2)}
\end{equation}
Assuming that the contribution to kSZ signal comes only from small scales $|\bb{l_2}| >> |\bb{l}|$, we can simplify this expression to get:
\begin{equation}
    \lambda^{-1}(l) \approx 2\pi\int dl_2l_2 \frac{(P_{\alpha}^{g\tau}(l_2))^2}{\tilde{P}^{\theta}(l_2)\tilde{P}^{g}_{\alpha}(l_2)}
\end{equation}
When calculating the integral for $\hat{v}_\alpha(\bb(l))$, we notice that weights neatly factorize: $ W_{\alpha}(\bb{l_1},\bb{l_2}) = W_\theta(\bb{l_1})W_{g\alpha}(\bb{l_2})$. 
We thus find the quadratic estimator of the radial velocity field\footnote{In \cite{Giri_2022} it was discussed that it is somewhat ambiguous what the true velocity field $v_r$ is, which $\hat{v}_r$ is supposed to reconstruct. In the case of N-body simulations, which are a collection of particles, one most directly obtains the sum over particles
\begin{align}
    \bb{q}(\bb{x}) = \frac{1}{n_p} \sum_i \bb{v}_i \delta^3(\bb{x}-\bb{x_i})
\end{align}
where $n_p$ is the number density average over the whole volume. This quantity is the particle or mass-weighted velocity field, i.e. the \emph{momentum field}. From the momentum field, one can define the \emph{velocity field} by choosing a smoothing scale and defining the velocity to be the smoothed momentum, divided by the smoothed density (appropriately regulated to avoid dividing by zero in voids). In \cite{Giri_2022} it was shown that the momentum field correlates somewhat better with the estimator $\hat{v}_r$ than the velocity field defined in this way. However, on the theoretical side, it is easier to work with velocities rather than momenta since they are first order in the perturbations. For this reason, in this work, we use the velocity field rather than the momentum field. On large scales, the difference between the two quantities is small, because the anisotropy in the radial momentum field $q_r = (1+\delta)v_r$ is dominated by the larger fluctuations in $v_r$.
}:
\begin{equation}
\label{eq:qe}
    \hat{v}_\alpha(\bb{l}) = (2\pi)^2\lambda(l)\int d^2x  [\theta_{obs}(\bb{x})]_{W_\theta(l)}[\delta_{obs}^{g,\alpha}(\bb{x})]_{W_g(l)}
\end{equation}
where 
\begin{align}
[f(\bb{x})]_{W(l)} \equiv \frac{1}{(2\pi)^2} \int d^2l f(\bb{l})W(l)e^{i\bb{l}\bb{x}};\hspace{1cm}  W_\theta(l) = \frac{1}{\tilde{P}^{\theta}(l)};\hspace{1cm}  W_{g,\alpha}(l) = \frac{P^{g\tau}_\alpha(l)}{\tilde{P}^{g}_\alpha(l)}    
\end{align}
The prefactor $(2\pi)^2\lambda(l)$ is the noise of the quadratic estimator. We can split the correlation function of QE velocities into two terms: $\langle\hat{v}_\alpha(\bb{l})\hat{v}_{\beta}(\bb{l'})\rangle = (N^{QE}+S^{QE})(2\pi)^2\delta^{2D}(\bb{l}+\bb{l'})$. Then the expression for noise is as follows:
\begin{equation}
N^{QE}=(2\pi)^2\left(\int d^2l_1d^2l_2 \delta^{2D}(\bb{l-l_1-l_2})\frac{(P_{\alpha}^{g\tau}(l_2))^2}{\tilde{P}^{\theta}(l_1)\tilde{P}^{g}_{\alpha}(l_2)}\right)^{-1}    
\end{equation}
The noise thus depends on the CMB noise $\theta(\bb{l})^{N}$ as well as the noise of the large-scale structure tracer. As a large-scale structure tracer, we use the halo (or galaxy) field $\delta^{g,\alpha}_{obs}(\bb{l}) = \delta^g_{\alpha}(\bb{l})+n^g(\bb{l})$. Here the stochastic shot noise term $n^g(\bb{l})$ comes from discrete sampling and has a power spectrum $N^g(l) = n(z)^{-1}$, where $n(z)$ is number of halos per redshift interval per steradian. In this derivation, we assumed that $\langle\tau_{\alpha}\delta^g_{\beta}\rangle \propto \delta^{(K)}_{\alpha,\beta}$, i.e., that small-scale matter fields are uncorrelated between different redshift bins. In \cite{Cayuso_2021} it was shown how to include redshift bin correlations of the small-scale fields in the QE. However, we found that for our study with 60 redshift bins, this correlation is small and we neglect it both in our QE implementation and our likelihood.

\section{Likelihood approach}
\label{sec:likeli}

We now discuss the different elements of our novel likelihood formulation. We discuss several different likelihoods with different properties, in increasing order of complexity.

\subsection{Likelihood for the velocity field from a kSZ  observation assuming a known matter field}
\label{sec:Lvonly}

We can write a maximum likelihood estimator for the velocities by writing down the Gaussian likelihood of CMB perturbations including the kSZ:
\begin{align}
\label{eq:Lkszvel}
- 2 \ln \mathcal{L} \left( \theta^{obs} | v_r \right)  =& \left( \theta^{obs} - \theta^{kSZ}(\delta_e,v_r) \right)^T P^{-1}_{\theta} 
\left( \theta^{obs} - \theta^{kSZ}(\delta_e,v_r) \right)  + \mathrm{const.}
\end{align}
Here $\theta^{obs} = \theta^{pCMB} + \theta^{kSZ} + n_\theta$ is the lensed primary CMB with added noise and $P = S + N$ is the covariance of $\theta^{pCMB}+n_\theta$ and $\theta^{pCMB}$ can also include uncorrelated kSZ such as from reionization (however assuming Gaussianity thereof). We use a basis-independent notation where $\theta$ are 2d fields and $\delta_e$ and $v_r$ are redshift binned 2d fields, both represented as a vector. 

The kSZ is given in position space by the radial integral 
\begin{align}
\label{eq:ksz}
\theta^{kSZ}(n) = f_\tau \int dr \ \delta_e(r,n) \ v_r(r,n)
\end{align}
and we can write it basis-independent in index notation as
\begin{align}
\theta^{kSZ}_l = K_l^{x,y} \delta_x v_y
\end{align}
where $K_l^{x,y}$ is the kSZ projection matrix.

This likelihood is the same as in \cite{arxiv.2205.15779}. Assuming that one has an estimate of $\delta_e$ for example from a galaxy survey (see below), the maximum likelihood estimator for the velocities can in principle be found analytically by evaluating 
\begin{align}
    \frac{\delta \mathcal{L}}{\delta v} \stackrel{!}{=} 0
\end{align}
However, in the case of multiple redshift bins, this estimator is ill-defined, and even for a single bin it gives a poor reconstruction in terms of residuals with the truth. In \cite{arxiv.2205.15779} this problem was circumvented by introducing a coarse-graining procedure. Here we follow a more Bayesian approach and include a Gaussian prior on the velocities, which is physically appropriate at large scales. The posterior is then 
\begin{align}
\label{eq:Pkszvel}
- 2 \ln \mathcal{P} \left( \theta^{obs} | v_r \right)  =& \left( \theta^{obs} - \theta^{kSZ}(\delta_e,v_r) \right)^T P^{-1}_{\theta} 
\left( \theta^{obs} - \theta^{kSZ}(\delta_e,v_r) \right) + v^T P_v^{-1} v + \mathrm{const.}
\end{align}
and the maximum a posteriori (MAP) is given by 
\begin{align}
    \frac{\delta \mathcal{P}}{\delta v} \stackrel{!}{=} 0
\end{align}
We will derive an analytic solution for this case in the next section, however, in practice we maximize the posterior numerically. The simple posterior discussed in this section has the advantage that there is only a single unknown field, the velocities $v$. Below we will jointly fit the matter and velocity fields as probabilistic degrees of freedom and we will not use this likelihood in our main implementation. However, in appendix \ref{app:errorbars} we use the posterior in Eq. \eqref{eq:Pkszvel} to demonstrate that we can obtain correct error bars from its Hessian.

\subsection{Analytic MAP estimator for the velocity field}
\label{sec:analyticMAP}
While ultimately we find MAP solutions numerically, it is interesting to consider whether there is an analytic MAP estimator. As we shall see, this is possible in the case of a known $\tau$ field (or one that has been reconstructed from galaxies independently of the kSZ data). The estimator we develop here is different from the estimator in \cite{arxiv.2205.15779} in that we include a Gaussian prior on the large-scale velocity field, thus obtaining a maximum a posteriori (rather than maximum likelihood) estimator.

We assume that we observe some $\theta_{o}$ which is sourced by $\theta_{p}$, primary CMB anisotropy, and $\theta_{kSZ}$ with Gaussian noise with known covariance $N_{\theta}$. In this section, we write the binned electron field in terms of $\tau$ rather than in terms of $\delta_e$ for compactness of notation. The two quantities are related by Eq.\eqref{eq:tau}. Following our discretization of the radial kSZ integral \eqref{eq:ksz_binned}, in position space we have $\theta_{kSZ} = \sum^{N_{bins}}_{i=1}v_i\tau_i$, with both $v_i$ and $\tau_i$ being 2D fields. We include Gaussian priors on velocities and primary CMB and assume that we know the optical depth $\tau$. Then, we can write a posterior:
\begin{equation}
    \mathcal{P}(v_i,\theta_p|\theta_o,\tau_i) \simeq e^{-\frac{1}{2}\left [ (\theta_{o}-\theta_{p}-\tau_iv^i)^TN_{\theta}^{-1}(\theta_{o}-\theta_{p}-\tau_iv^i)+(v^T)^i(C^{-1}_v)_{ij}v^j+\theta_p^TC^{-1}_{\theta}\theta_p\right ]}
\end{equation}
Weak equality $\simeq$ here means equality up to a constant, multiplicative normalization (or additive constant if $\mathrm{log}$-quantities are considered) of $\mathcal{N}^{-1} = \sqrt{\det C_{\theta}}\times\sqrt{ \det C_v }\times\sqrt{\det C_\theta}$. In the expression of a type $\theta^T_pN_{\theta}^{-1}\theta_p$, a sum over some discretization (pixels) of a field is implied, i.e. $\theta^TN_{\theta}^{-1}\theta \equiv \sum^{N_{pix}}_{p_i,p_j}(\theta)_{p_i}(\theta)_{p_j}(N_{\theta}^{-1})_{p_i,p_j}$. We are interested in the MAP for $v_i$, hence we are going to neglect the normalization (this is possible since $v_i$ does not appear in covariance) and marginalize over the primary CMB $\theta_p$:
\begin{equation}
    \mathcal{P}(v_i|\theta_o,\tau_i) = \int D\theta_p \mathcal{P}(v_i,\theta_p|\theta_o,\tau_i) 
\end{equation}
Let's consider the $\theta_p$ part:
\begin{equation}
    (-2\ln\mathcal{P}(v_i,\theta_p|\theta_o,\tau_i))_{\theta_p} \simeq \left [ \theta^T_pC^{-1}_{\theta}\theta_p + \theta_p^TN^{-1}_{\theta}\theta_p - (\theta_o-\tau_iv^i)^TN_\theta^{-1}\theta_p - \theta^T_pN_\theta^{-1}(\theta_o-\tau_iv^i) \right ]
\end{equation}
We define $\tilde{C}^{-1}_{\theta}\equiv C^{-1}_\theta + N^{-1}_\theta$ and $\tilde{\theta} \equiv \tilde{C}_\theta N^{-1}(\theta_o-\tau_iv^i)$. Then we can rewrite:

\begin{equation}
    (-2\ln\mathcal{P}(v_i),\theta_p|\theta_o,\tau_i))_{\theta_p} \simeq \theta_p^T\tilde{C}^{-1}_{\theta}\theta_p - \tilde{\theta}^T\tilde{C}^{-1}_{\theta}\theta_p -  \theta^T_p\tilde{C}^{-1}_\theta\tilde{\theta} = (\theta_p-\tilde{\theta})^T\tilde{C}^{-1}_{\theta}(\theta_p-\tilde{\theta}) - \tilde{\theta}^T\tilde{C}^{-1}_{\theta}\tilde{\theta}
\end{equation}
Now we can perform marginalization over $\theta_p$ easily. Since $\int D\theta_p e^{-\frac{1}{2}(\theta_p-\tilde{\theta})^T\tilde{C}^{-1}_{\theta}(\theta_p-\tilde{\theta})} \simeq 1$, we are left with

\begin{equation}
    -2\ln\mathcal{P}(v_i|\theta_o,\tau_i) \simeq  (\theta_o-\tau_iv^i)^TN_\theta^{-1}(\theta_o-\tau_iv^i) - (\theta_o-\tau_iv^i)^TN_\theta^{-1}\tilde{C}_{\theta}N_\theta^{-1}(\theta_o-\tau_iv^i) + (v^T)^i(C^{-1}_v)_{ij}v^j 
\end{equation}
We can try to rewrite this expression as we did before for $\theta_p$. The velocity part is as follows:
\begin{align}
    (-2\ln\mathcal{P}(v_i|\theta_o,\tau_i))_{v} & \simeq
    (v^T)^i((C^{-1}_v)_{ij}+(\tau^T)_i(N^{-1}_{\theta}-N^{-1}_{\theta}\tilde{C}_\theta N^{-1}_{\theta})\tau_j)v^j \\ &-(v^T)^i(\tau^T)_i(N^{-1}_{\theta}-N^{-1}_{\theta}\tilde{C}_\theta N^{-1}_{\theta})\theta_o - \theta^T_o(N^{-1}_{\theta}-N^{-1}_{\theta}\tilde{C}_\theta N^{-1}_{\theta})\tau_jv^j
\end{align}
Assuming that the inverse exists, we can define a $\tau$ dependent covariance: 
\begin{equation}
\label{eqn:cov_v}
(\tilde{C}^{-1}_v)_{ij} \equiv (C^{-1}_v)_{ij}+(\tau^T)_i(N^{-1}_{\theta}-N^{-1}_{\theta}\tilde{C}_\theta N^{-1}_{\theta})\tau_j
\end{equation}
Then $\tilde{v_i}$, which is the MAP estimator we were looking for, is given by:
\begin{equation}
\label{eq:mapestimator}
\tilde{v}_i = (\tilde{C_v})_i^j(\tau^T)_j(N^{-1}_{\theta}-N^{-1}_{\theta}\tilde{C}_{\theta}N^{-1}_{\theta})\theta_o
\end{equation}
Now the log-likelihood takes the following form:
\begin{equation}
    -2\ln\mathcal{P}(v_i|\theta_o,\tau_i) \simeq (v^T-\tilde{v}^T)^i(\tilde{C}^{-1}_v)_{ij}(v-\tilde{v})_j - 
    (\tilde{v}^T)^i(\tilde{C}^{-1}_v)_{ij}(\tilde{v})_j + \theta^T_o(N^{-1}_{\theta}-N^{-1}_{\theta}\tilde{C}_\theta N^{-1}_{\theta})\theta_o
\end{equation}
It is worth noting that had we instead treated $\tau_i$ as unknown, we could have added $-2\ln\mathcal{P}(\tau_i|\tau_o) = (\tau_o^T-\tau^T)_{i}(N_\tau^{-1})^{ij}(\tau_o-\tau)_{j} + (\tau^T)_{i}(C_\tau^{-1})^{ij}(\tau)_j$ and, after marginalizing over $\theta_p$, marginalized over $\tau$. That procedure would result in a posterior for velocity $\mathcal{P}(v_i|\theta_o,\tau_o)$ that depends only on observed data. However, finding an analytical MAP would not be possible in that case, which is consistent with the fact that it is not possible to further analytically marginalize the posterior that we obtained over $\tau_i$. We also note that our MAP depends on an inverse of the operator that is explicitly $\tau_i$-dependent. It is worth adding that one way to proceed with an analytic MAP for the velocity, in this case, is to find the MAP of $\tau_i$ alone from galaxy survey data and use those values to evaluate the covariance operator the Eq. \eqref{eqn:cov_v}.

Our result is consistent with the results of \cite{arxiv.2205.15779}, where the inverse was computed via the "coarse-graining" procedure. Indeed, Eq. (14) in \cite{arxiv.2205.15779}, which is the MLE condition there, reads as follows:
\begin{equation}
    (\tau^\alpha)^\dag(C^{pCMB})^{-1}\Theta =  (\tau^\alpha)^\dag(C^{pCMB})^{-1}(\tau^\beta)v_{\beta}
\end{equation}
That's equivalent to our case if we neglect the prior and consider that we observe $\theta^p$. As was mentioned in the same work, QE is obtained if we approximate $(\tau^\alpha)^\dag(C^{pCMB})^{-1}(\tau^\beta)$ with $\langle (\tau^\alpha)^\dag(C^{pCMB})^{-1}(\tau^\beta) \rangle_{\tau}$. In that case, the covariance is independent of $\tau$ and the resulting expression for MAP $v$ is quadratic in the observed fields.

Our analytic MAP estimator given in Eq. \eqref{eq:mapestimator} is not easy to evaluate in practice, because it includes several inverses of large matrices. For this reason, even for the velocity field likelihood presented in this section, we find the MAP by gradient descent rather than from the analytic expression.

\subsection{Joint likelihood for CMB and matter}

We now describe a likelihood where the fields $v,\delta_m,\theta^{CMB}$ are treated probabilistically. In this section, we assume that $\delta_e=\delta_m$ and that we can directly observe $\delta_m$ with some Gaussian noise (which we will take to have a flat shot noise power spectrum). We also assume that both the galaxy survey and the CMB experiment have mutually uncorrelated Gaussian noise. The likelihood is then given by
\begin{align}
\label{eq:Lkszmatter}
- 2 \ln \mathcal{L} \left( \theta^{obs},\delta_{m}^{obs} | \theta^{pCMB},\delta_m,v_r \right)  =& \left( \theta^{obs} - \theta^{pCMB} - \theta^{kSZ}(\delta_m,v_r) \right)^T N^{-1}_{\theta} \nonumber \\
& \times \left( \theta^{obs} - \theta^{pCMB} - \theta^{kSZ}(\delta_m,v_r) \right)  \nonumber \\
  & + \left( \delta_m^{obs} - \delta_m \right)^T N^{-1}_m \left( \delta_m^{obs} - \delta_m \right) + \mathrm{const.}
\end{align}
We again use a basis-independent notation where $\theta$ is a 2d field and $\delta_m$ and $v_r$ are redshift binned 2d fields, both represented as a vector. 

To get the maximum a posteriori estimate (MAP) we maximize the product of the likelihood times priors for the unobserved quantities. By Bayes theorem:
\begin{align}
\label{eq:posteriorkszmatter}
 \mathcal{P} \left( \theta^{CMB},\delta_m,v_r | \theta^{obs},\delta_{m}^{obs} \right)  \propto & \ \mathcal{L} \left( \theta^{obs},\delta_{m}^{obs} | \theta^{CMB},\delta_m,v_r \right) \times \mathcal{P}(T^{CMB},\delta_m,v_r)
\end{align}
If we assume Gaussian priors for the fields we get
\begin{align}
 -2 \ln \mathcal{P}(\theta^{CMB},\delta_m,v_r) = (\theta^{CMB})^T P_\theta^{-1} \theta^{CMB} + v^T P_v^{-1} v + \delta_m^T P_{m}^{-1} \delta_m + \mathrm{const.}
\end{align}
where $P_x$ are the signal covariance matrices (power spectra) for fiducial cosmological parameters. In this prior we have assumed a separation of scales, i.e. we draw the velocities $v_r^L$ from a Gaussian field on large scales, and we draw the matter field $\delta_m^S$ on small scales only. In our implementation, we split these scales at $\ell_{\mathrm{split}}=700$. The analysis is not sensitive to the precise choice of $\ell_{\mathrm{split}}$
    because kSZ is induced by significantly smaller-scale fluctuations of matter overdensity that are modulated by substantially larger-scale velocity fluctuations. In addition, we can only observe kSZ at much higher $\ell$ (where it becomes a dominant anisotropy) and we can only reconstruct velocities at much lower $\ell$ (due to signal-to-noise). Our likelihood thus treats the velocities on large scales $v^L$ independently from the matter field on small scales $\delta_m^S$. This neglects the gravitation induced correlation between $v^L$ and $\delta^S$, such as the bispectrum of form $\left< v^L \delta_m^S \delta_m^S \right>$. However, these squeezed limit correlations are small and can be neglected. For some applications of parameter fitting, one would also like to include information from the large-scale galaxy field and enforce a relation between matter density and velocity (i.e. the continuity equation on large scales). However, for our present purpose of reconstructing the large-scale velocity field, such an extended likelihood is not required.

We now want to maximize the posterior with respect to $v_r, \delta_m, \theta^{CMB}$. Unlike the case in Sec. \ref{sec:analyticMAP}, there is no analytic solution to the MAP. 
We thus use optimization to find the MAP of the fields. In particular, $\delta_m$ is now constrained from both the observed matter data and the observed kSZ data (for example, a cluster is visible both to a galaxy survey and a high-resolution CMB experiment, so they contain mutual information). Our likelihood formulation is optimal if the noise covariances $N$ and signal covariances (in the prior) are correct, including bin-to-bin correlations in the signal. This assumes Gaussianity in all fields. We will discuss how to relax this assumption with machine learning below.

\subsection{Joint likelihood for CMB and galaxies}
\label{jointlikeli}

In realistic experiments, our data sets are the observed CMB temperature anisotropy $\theta^{obs}$ and the observed galaxy overdensity $\delta^{obs}_g$ (rather than $\delta_m$ or $\delta_e$). We assume that these quantities are determined by the following underlying unobserved fields: true radial velocity $v_r$, true galaxy overdensity $\delta_g$ (without stochastic shot noise), true electron density $\delta_e$, and true lensed primary CMB temperature $\theta^{pCMB}$. We also assume that both the galaxy survey and the CMB experiment have mutually uncorrelated Gaussian noise. In this case the log-likelihood of seeing $\theta^{obs}$ and $\delta^{obs}_g$ for given $v_r, \delta_g, \delta_e, \theta^{pCMB}$ is the following:
\begin{align}
\label{eq:likelinoise}
- 2 \ln \mathcal{L} \left( \theta^{obs},\delta_{g}^{obs} \right)  =& \left( \theta^{obs} - \theta^{pCMB} - \theta^{kSZ}(\delta_e,v_r) \right)^T N^{-1}_{T} \left( \theta^{obs} - \theta^{pCMB} - \theta^{kSZ}(\delta_e,v_r) \right)  \nonumber \\
  & + \left( \delta_g^{obs} - \delta_g \right)^T N^{-1}_g \left( \delta_g^{obs} - \delta_g \right) + \mathrm{const.}
\end{align}
where $N_{T}$ is the noise covariance of the CMB experiment and $N_g$ is the shot noise of the galaxy survey due to the finite number of observed galaxies. A crucial question is now how to relate the galaxy density $\delta_g$ and the unobserved electron density $\delta_e$. There are two general options how to accomplish this, deterministic or stochastic, as we now discuss. In our numerical tests below we will only implement the deterministic approach.

In the expression above we have assumed that the noise of the CMB field and the noise of the galaxy field are uncorrelated. This holds exactly true in our main analysis (Sec. \ref{sec:resdeltam}), because we approximate the galaxy noise by a constant pixel-independent shot noise, which is not correlated with the primary CMB, CMB instrumental noise, or kSZ. In reality, however, a part of the kSZ, as well as foreground residuals which we do not model, come from the halos that the galaxy survey resolves, which leads to a correlation between CMB and galaxy noise residuals. 
To take into account a noise correlation we define $n_t = \theta^{obs} - \theta^{pCMB} - \theta^{kSZ}(\delta_e, v_r)$ and $n_g = \delta_g^{obs} - \delta_g$. Then we can define a vector $n^T = (n_t,n_g)$ and a covariance matrix $C_N$:
\begin{equation}
    C_N = \begin{pmatrix}
N_T & \mathrm{Cov}(n_t, n_g) \\
\mathrm{Cov}(n_g, n_t) & N_g
\end{pmatrix}
\end{equation}
Then the likelihood equation takes the usual correlated Gaussian form:
\begin{equation}
- 2 \ln \mathcal{L} \left( \theta^{obs},\delta_{g}^{obs} | \theta^{pCMB},\delta_g,\delta_e,v_r \right) = n^T (C_N)^{-1} n + \mathrm{const.}
\end{equation}
The correlated noise $\mathrm{Cov}(n_t, n_g)$ could be estimated from simulations or perhaps be calculated in the halo model. We also note that the noise PDF on small scales is not guaranteed to be Gaussian, and could be learned with a machine learning approach. In the following, we do not take into account noise correlation and instead work with Eq. \ref{eq:likelinoise}, which, as we discussed, holds exactly for our Poisson noise setup (Sec. \ref{sec:resdeltam}), and also performs well for the halo analysis (Sec. \ref{eq:resultsdeltah}).

\subsubsection{Using an estimator for $\delta_e$ given $\delta_g$}
\label{sec:dedgtemplate}

The most straight forward way to estimate $\delta_e$ from $\delta_g$ is the following template estimator (e.g. \cite{arxiv.2205.15779,arxiv.1810.13423}):
\begin{equation}
\label{eq:electrontemplate}
\hat{\delta}_e(k) = \frac{P_{eg}}{P_{gg}}\delta_g(k)
\end{equation}
where the small-scale power spectra $P_{eg}$ and $P_{gg}$ can be calculated in the halo model (see \cite{arxiv.1810.13423}) or estimated from simulations (which we do here).

We can then write the likelihood as
\begin{align}
\label{eq:Tgallikeli}
- 2 \ln \mathcal{L} \left( \Theta^{obs},\delta_{g}^{obs} | \theta^{pCMB},\delta_g,v_r \right)  =& \left( \theta^{obs} - \theta^{pCMB} - \theta^{kSZ}(\hat{\delta}_e(\delta_g),v_r) \right)^T N^{-1}_{T} \nonumber \\
& \times \left( \theta^{obs} - \theta^{pCMB} - \theta^{kSZ}(\hat{\delta}_e(\delta_g),v_r) \right)  \nonumber \\
  & + \left( \delta_g^{obs} - \delta_g \right)^T N^{-1}_g \left( \delta_g^{obs} - \delta_g \right) + \mathrm{const.}
\end{align}
The kSZ calculated from $\hat{\delta}_e$ will only include the resolved kSZ, i.e. the one that can be explained by the galaxy distribution. The estimator in Eq. \eqref{eq:electrontemplate} is not guaranteed to be the optimal estimator since these small-scale fields are non-Gaussian. Below in Sec. \ref{sec:machinelearning} we will explore a machine learning method, using a Convolutional Neural Network to learn a better template $\hat{\delta}_e(\delta_g)$. If we assume Gaussianity for the fields we also get the prior
\begin{align}
\label{eq:Tgalposterior}
 -2 \ln \mathcal{P}(\theta^{pCMB},\delta_g,v_r) = (\theta^{pCMB})^T P_T^{-1} \theta^{pCMB} + v^T P_v^{-1} v + \delta_g^T P_{g}^{-1} \delta_g + \mathrm{const.}
\end{align}
where $P_x$ are the signal covariance matrices for fiducial cosmological parameters.

\subsubsection{Coupling $\delta_e$ and $\delta_g$ with a correlated prior}

A second, more Bayesian, option is to impose a stochastic relation between $\delta_e$ and $\delta_g$. In this case, the likelihood is
\begin{align}
- 2 \ln \mathcal{L} \left( \theta^{obs},\delta_{g}^{obs} | \theta^{pCMB},\delta_g,\delta_e,v_r \right)  =& \left( \theta^{obs} - \theta^{pCMB} - \theta^{kSZ}(\delta_e,v_r) \right)^T N^{-1}_{T} \nonumber \\
 & \times \left( \theta^{obs} - \theta^{pCMB} - \theta^{kSZ}(\delta_e,v_r) \right)  \nonumber \\
  & + \left( \delta_g^{obs} - \delta_g \right)^T N^{-1}_g \left( \delta_g^{obs} - \delta_g \right) + \mathrm{const.}
\end{align}
To impose a relation between $\delta_e$ and $\delta_g$, we need a correlated prior. If we assume the Gaussianity of the small-scale fields, the prior is 
\begin{align}
\label{eq:Pgegauss}
 -2 \ln \mathcal{P}(\delta_e,\delta_g) = 
\begin{pmatrix} \delta_e & \delta_g  \end{pmatrix} \begin{pmatrix} P_{ee} & P_{ge} \\ P_{ge} & P_{gg} \end{pmatrix}^{-1} \begin{pmatrix}  \delta_e \\  \delta_g \end{pmatrix} + \mathrm{const.}
\end{align}
where the covariance matrix must be invertible, i.e. the two fields cannot be fully correlated (as is the case if one naively assumes $\delta_g \propto \delta_e$). The posterior is then
\begin{align}
 \mathcal{P} \left( \theta^{pCMB},\delta_g,v_r,\delta_e | T^{obs},\delta_{g}^{obs} \right)  \propto & \ \mathcal{L} \left( T^{obs},\delta_{g}^{obs} | T^{pCMB},\delta_g, \delta_e,v_r \right) \times \mathcal{P}(T^{pCMB},\delta_e, \delta_g,v_r)
\end{align}
In this approach, again the assumption of Gaussianity could be circumvented by machine learning. A non-Gaussian prior $\mathcal{P}(\delta_e,\delta_g)$ can in principle be learned from simulations using the technique of normalizing flows \cite{Rouhiainen:2022cwd,Rouhiainen:2021qjg}. We are planning to explore this idea in the future.

\subsection{Outlook: Jointly estimating cosmological and astrophysical parameters}
\label{sec:parameters}

So far we have discussed how to estimate fields, assuming that the priors are known, i.e. they have been evaluated for known astrophysical and cosmological parameters. We now generalize the posterior to include such parameters. We are splitting this discussion into two types of parameters, according to our separation of scales. We defer an application of such a fit to simulations to future work. 

Astrophysical parameters affect the small-scale $\delta_e$ and $\delta_g$ fields, in the deeply non-linear regime where most of the observed kSZ signal is generated. For example, one can calculate $P_{ge}$ in the halo model as a function of the slope of the radial electron profile (see \cite{arxiv.1810.13423}). $P_{ge}$ appears in the likelihood above in Eq. \eqref{eq:electrontemplate} and Eq. \eqref{eq:Pgegauss}. For example, if we assume Gaussian fields, the prior in Eq. \eqref{eq:Pgegauss}, written in terms of the vector $\delta_{ge}=\begin{pmatrix}  \delta_e \\  \delta_g \end{pmatrix}$ becomes
\begin{align}
 -2 \ln \mathcal{P}(\delta_e,\delta_g|\Lambda_S) = 
\delta_{ge}^T C_{ge}^{-1}(\Lambda_S) \delta_{ge} + \log \det C_{ge}(\Lambda_S)  
\end{align}
Note that now we need to keep track of the determinant factor since it depends on the $\Lambda_S$ parameters which we optimize for. We can use a prior on the astrophysical parameter $\Lambda_S$ to express $\mathcal{P}(\delta_e,\delta_g,\Lambda_S) = \mathcal{P}(\delta_e,\delta_g,|\Lambda_S) \mathcal{P}(\Lambda_S)$

Cosmological parameters affect $v_r$ and $\delta_g$ at large scales. To avoid complications due to non-linear evolution (which could be included with a differentiable forward model of structure formation) we can probe cosmological parameters on linear scales $k \ll k_{NL}$. In particular, we can constrain local non-Gaussianity $f_{NL}$ \cite{M_nchmeyer_2019}, galaxy bias $b_g$, kSZ optical depth bias $b_v$ \cite{arxiv.2205.15779} and the Hubble constant $H$. To include constraining power from the galaxy field on large scales, we need to add a field likelihood for the large-scale galaxy field $\delta_{g,L}$ given by
\begin{align}
- 2 \ln \mathcal{L} \left(\delta_{g,L}^{obs} | \delta_{g,L} \right) = \left( \delta_{g,L}^{obs} - \delta_{g,L} \right)^T N^{-1}_g \left( \delta_{g,L}^{obs} - \delta_{g,L} \right)    
\end{align}
The joint prior $\mathcal{P}(\delta_{g,L},v_r)$ of the fields $\delta_{g,L}$ and $v_r$ on large (linear) scales as a function of cosmological parameters $\lambda_L$ is Gaussian. Written in terms of the vector $\delta_{gv}=\begin{pmatrix}  \delta_{g,L} \\  v_r \end{pmatrix}$ it is given by 
\begin{align}
\label{eq:Pgvgauss}
 -2 \ln \mathcal{P}(\delta_{g,L},v_r|\Lambda_L) = 
\delta_{gv}^T C_{gv}^{-1}(\Lambda_L) \delta_{gv} + \log \det C_{gv}(\Lambda_L)  
\end{align}
Analytic expressions for the covariance matrix $C_{gv}$ on linear scales are given for example in \cite{M_nchmeyer_2019}.

Including all these parameters, in full generality, the posterior is 
\begin{align}
 \mathcal{P} \left( \theta^{pCMB},\delta_g,\delta_e,v_r, \Lambda_S, \Lambda_L | \theta^{obs},\delta_{g}^{obs} \right)  \propto & \ \mathcal{L} \left( \theta^{obs},\delta_{g}^{obs} | \theta^{pCMB},\delta_g, \delta_e,v_r \right) \times \mathcal{P}(\theta^{pCMB},\delta_e, \delta_g,v_r, \Lambda_S, \Lambda_L)
\end{align}
Physical constraints are then various joint or marginal distributions of this complicated posterior. Finding the MAP and marginalizing over fields will be challenging and is deferred to future work. 

\subsection{Implementation and optimization of the likelihood}
\label{sec:implementation}

We implement the likelihood equations in JAX - a Python package with a versatile and highly optimized autodifferentiation. In this way, we can take analytic derivatives with respect to all fields and scalar parameters. We then find the maximum a posteriori (MAP), i.e. the parameter and field configuration which maximizes the posterior probability density. 
We now give a detailed description of our python implementation, which is based on the JAX library.

\textbf{Data dimensionality.} Given $n$ redshift bins and for an angular side length $N_{\mathrm{side}}$\footnote{$N_{side}$ here represents the number of pixels per side of a square patch. In our simulation results presented below, it also coincides with the $N_{side}$ parameter for the input healpix maps, since the total number of pixels in a healpix map with given $N_{side}$ is $N_{pix}=12\times N^2_{side}$ and we analyze exactly $1/12$ of the fullsky healpix map, projected to a flat patch.}, we have in total $N_{\mathrm{side}}^2\times(2n+1)$ parameters to solve for ($n$ velocities, $n$ electron densities, and the primary CMB).

\textbf{Masking the data.} Since we have non-periodic simulation data, we need to apodize the data with a smooth mask at the boundary. An exact treatment would deal with the mask by Wiener filtering with an inhomogeneous noise covariance matrix. However Wiener filtering at high signal-to-noise converges very slowly, so instead we use the common but somewhat suboptimal apodization technique (as implemented e.g. in \cite{Alonso:2018jzx}).

\textbf{Data representation.} The CMB likelihood term (e.g. Eq. \eqref{eq:Lkszmatter}) is naturally represented in pixel space ($x$-space) because the kSZ is obtained with a simple pixelwise multiplication and the noise covariance is diagonal in pixel space (even for inhomogeneous uncorrelated noise). The prior terms, however, are naturally represented in Fourier space ($l$-space), since covariance matrices are diagonal there. We thus use both representations and FFT between the two.

\textbf{Optimizer.}  We minimize the negative log posterior numerically for parameters of interest, for example, we minimize Eq. \eqref{eq:posteriorkszmatter} with respect to $\{\delta_m, v_r, \theta^{pCMB}\}$. We freely switch from $x$ to $l$-space and vice versa utilizing JAX FFT routines. We found that the popular Adam optimization algorithm, which is widely used for neural network training, works well for our setup. We used the implementation of Adam from the Optax library. Convergence of the optimization depends on the noise levels and required some tuning. We found that convergence can be improved by setting different initial learning rates for each field (for example $\sim 10^{-3}$ for velocity, $\sim 10^{-2}$ for density, and $\sim 10^{-5}$ for primary CMB anisotropy worked in many cases). We also used the exponential decline of learning rates during the optimization (using a decay factor of 0.83 every 150 optimization steps).

\textbf{Initialization.} We initialize $\{\delta_g = \delta^{obs}_g, v_r = \hat{v}^{QE}_r, \theta^{pCMB} = \theta^{obs}\}$. Here, $\hat{v}^{QE}_r$ is the quadratic estimator for $v_r$. Although all the results in this paper were obtained with $v_r = \hat{v}^{QE}_r$ initialization, we noticed that this is not essential and one can start with $v_r = 0$  as well.

\textbf{Multi-GPU parallelization.} Because of GPU memory limitations, there is a trade-off between resolution (both pixel-wise and bin-wise) and physical volume. However, we also found it possible to make use of several GPU devices in parallel to increase the size of the redshift interval of our analysis. We kept different groups of redshift bins on different GPUs and calculated kSZ temperature anisotropy after each optimization step via collective communication.

\textbf{Hardware and convergence time.} We can include up to 23 redshift-bins with maps of $2048^2$ pixels on a single 16 GB RTX-A4000. We also note, that in addition to parameters, one also needs to keep various constants in memory, such as the observed fields $\delta^{obs}_g, \theta^{obs}$, and theoretical covariances of the parameters. The optimization process on one GPU takes of order ten minutes. With the help of three GPUs we are able to include exactly three times more bins. For three GPUs the optimization time increased to approximately one hour, which includes inter-GPU and GPU-CPU communication. In our main analysis presented in Sec. \ref{sec:results}, we use 60 redshift bins spread over three GPUs.

\subsection{Joint posterior, marginal posterior, and error bars}
\label{subsubsec:errors}
The full posterior, including cosmological and astrophysical parameters which we keep fixed in our results below, is given by
\begin{align}
    \mathcal{P} \left( \theta^{pCMB},\delta_g,\delta_e,v_r, \Lambda_S, \Lambda_L | \theta^{obs},\delta_{g}^{obs} \right)
\end{align}
We will refer to this posterior as the \emph{joint posterior} (following the naming of lensing posteriors in \cite{Millea:2020cpw}). From this posterior one can, at least in principle, obtain various \emph{marginalized posteriors}. In particular, if one is only interested in reconstructing the velocity field, one may want to marginalize over the small-scale fields, i.e.
\begin{align}
    \mathcal{P} \left( v_r, \Lambda_S, \Lambda_L | \theta^{obs},\delta_{g}^{obs} \right) = \int d\theta^{pCMB} \ d \delta_g \ d \delta_e \ \mathcal{P} \left( \theta^{pCMB},\delta_g,\delta_e,v_r, \Lambda_S, \Lambda_L | \theta^{obs},\delta_{g}^{obs} \right)
\end{align}
In the CMB lensing likelihood analysis, references \cite{Hirata:2002jy,Carron:2017mqf} maximize the marginal posterior, while \cite{Millea:2017fyd,Millea:2020cpw} maximize the joint posterior. In the lensing case it is possible to analytically marginalize over the CMB fields, but maximizing the marginal posterior comes at the cost of repeatedly evaluating a computationally expensive determinant. In our present work, we maximize the joint posterior of the fields, but physically both the joint and marginal posteriors are interesting. To make unbiased estimates of cosmological or astrophysical parameters, it is in principle necessary to marginalize over the field variables. How this can be achieved in the case of CMB lensing was recently shown in \cite{Millea:2021had}, and we defer the kSZ case to future work.

In addition to finding the MAP we need to give error bars for the modes of the reconstructed velocity field $v_r$ (and ultimately also for cosmological and astrophysical parameters $\Lambda_S$ and $\Lambda_L$). This can be done by considering the curvature of the posterior (or, at least in principle, by sampling from the posterior). Assuming that the posterior is not multi-modal it can be expanded around the MAP to second order and error bars can be given in terms of the second derivatives of the posterior. As usual, we define the Hessian of the posterior
\begin{align}
    F_{ij} = -\frac{\partial^2 \ln \mathcal{P}}{\partial \lambda_i \partial \lambda_j} \Bigg|_{\lambda=\hat{\lambda}}
\end{align}
where $\lambda$ are all parameters of the posterior and $\hat{\lambda}$ is their MAP. One can obtain their un-marginalized error bars from the diagonal part as $\sigma_i = \frac{1}{\sqrt{F_{ii}}}$ and the marginalized ones as $\sigma_i = \left(F^{-1}\right)_{ii}$
if this inversion is tractable. Unfortunately inverting the full Hessian for the high-dimensional posterior with all fields is not computationally tractable. Here we restrict ourselves to a simpler setup. In App. \ref{app:errorbars} we show that for the simplest likelihood, given in Eq. \eqref{eq:Pkszvel}, we can obtain error bars on the reconstructed velocity field from the Hessian of the posterior, which match the residuals with the truth (which are available in simulations but not in real data). For this simpler likelihood, it is even possible to invert the full Hessian. At least in this simple case, we can thus set error bars on the reconstructed velocity field without the need to run large numbers of simulations, by using auto-differentiation to calculate the likelihood curvature at the MAP. We will investigate error bars in the full likelihood formulation, including methods to integrate out the small-scale fields either analytically or numerically, in future work.

\section{Application to simulations}
\label{sec:results}

In this section, we apply our likelihoods to a simulation. We explore under which conditions the MAP outperforms the QE and find that both a low CMB noise and a low large-scale structure noise are required for a significant improvement factor. 

\subsection{Agora simulation}

There are currently not many available simulations that have high-resolution halos, light-cone coordinates, include CMB, and cover a large sky fraction. The simulation requirements for kSZ velocity reconstruction are particularly challenging because one both needs the large velocity scales, and the small scales where the kSZ is visible, and here we also require a high halo density. One suitable simulation is the recently published Agora simulation, which we use here. Agora \cite{omori2022agora,Klypin_2016} is a multi-component simulation on a light cone that uses halos and particles from Multidark-Planck2 (MDPL2) N-body simulation and also models CMB primary and secondary anisotropies including lensing, kSZ, tSZ, and CIB. While here we only include the kSZ effect, other secondary anisotropies may be helpful in the future. 

With our hardware setup of three 16GB memory RTX-A4000 GPUs, we found the following resolution to be tractable. We downgrade the original full-sky maps from healpix $N_{side} = 8192$ to $N_{side} = 2048$, the latter corresponds to the resolution of $1.72\ \mathrm{arcmin}$. Then apply flat-sky approximation, keeping $1/12$ of the maps and treating them as fields on 2D Cartesian grid of resolution $2048^2$ pixels (see footnote 2 above). For this purpose, we found the \texttt{reproject} function from the \texttt{pixell} library useful. We consider the redshift interval of $z=0.5-3$ and use $N=60$ equidistant comoving radial bins so that each bin has a width of 50 Mpc. This binning merges the original 120 bins of Agora into pairs of two. Our optimization for 3-dimensional fields thus uses arrays of size $[60,2048,2048]$. In the future, when we apply our method to real data (CMB-S4), it would be useful to raise the angular resolution further, which would require a larger bank of GPUs. For our present goal of testing our likelihood method, the current resolution is sufficient.

\subsection{Velocity field, galaxy density and CMB map}
\label{sec:experimentdata}

\begin{figure}
\centering
\includegraphics[width=0.9\textwidth]{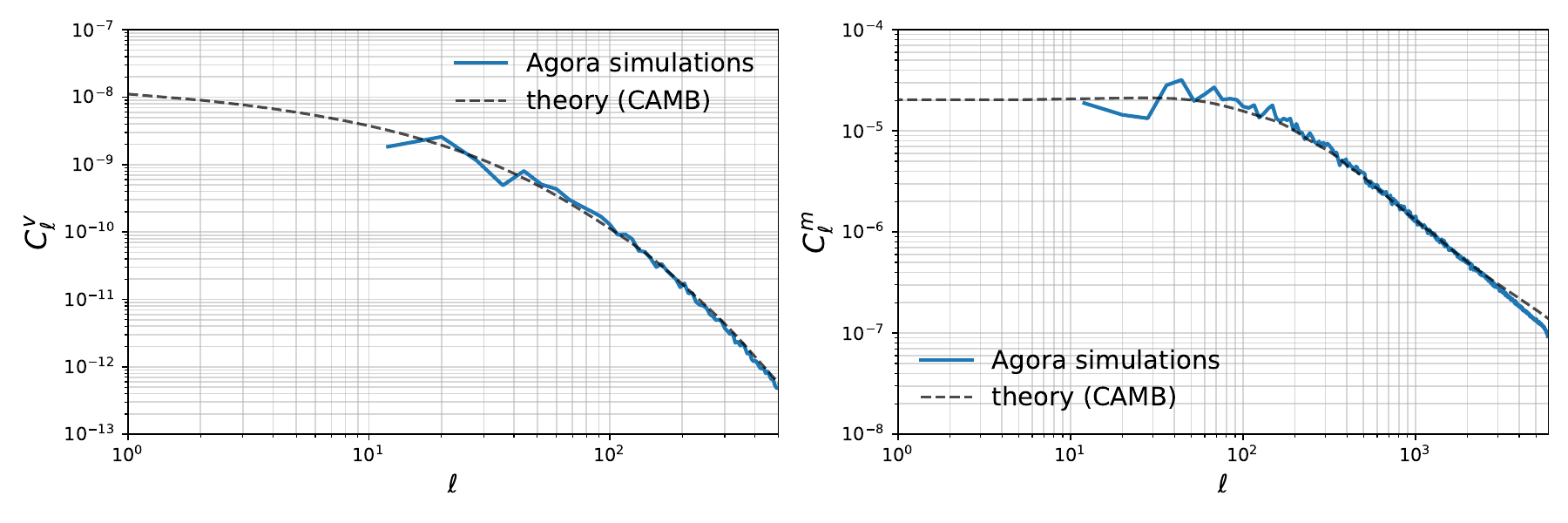}
\caption{Example of power spectra of radial velocity (left) and matter density (right) from Agora simulations (solid blue) compared to theoretical values calculated with CAMB (dashed black) at $z=1.15$. The power spectra were estimated from the apodized flatsky projection of 1/12 of the fullsky AGORA simulation.} \label{fig:vel_mat_ps}
\end{figure}

The first required simulation product in our analysis is the radial velocity field. Since the Agora simulation is patched together by repeating the same base simulation several times, this could potentially lead to wrong powers on large scales. In Figure \ref{fig:vel_mat_ps} we show the radial velocity field and the matter field of the Agora simulation for the radial bin at $z=1.15$, compared to the expectation from CAMB. The power spectra here and below were estimated from the simulation after taking a 1/12th of the sky flatsky projection and apodizing the mask. We find generally good agreement with theory at all redshifts, sufficient for our purpose.

\begin{figure}[t!]
\centering
\includegraphics[width=0.45\textwidth]{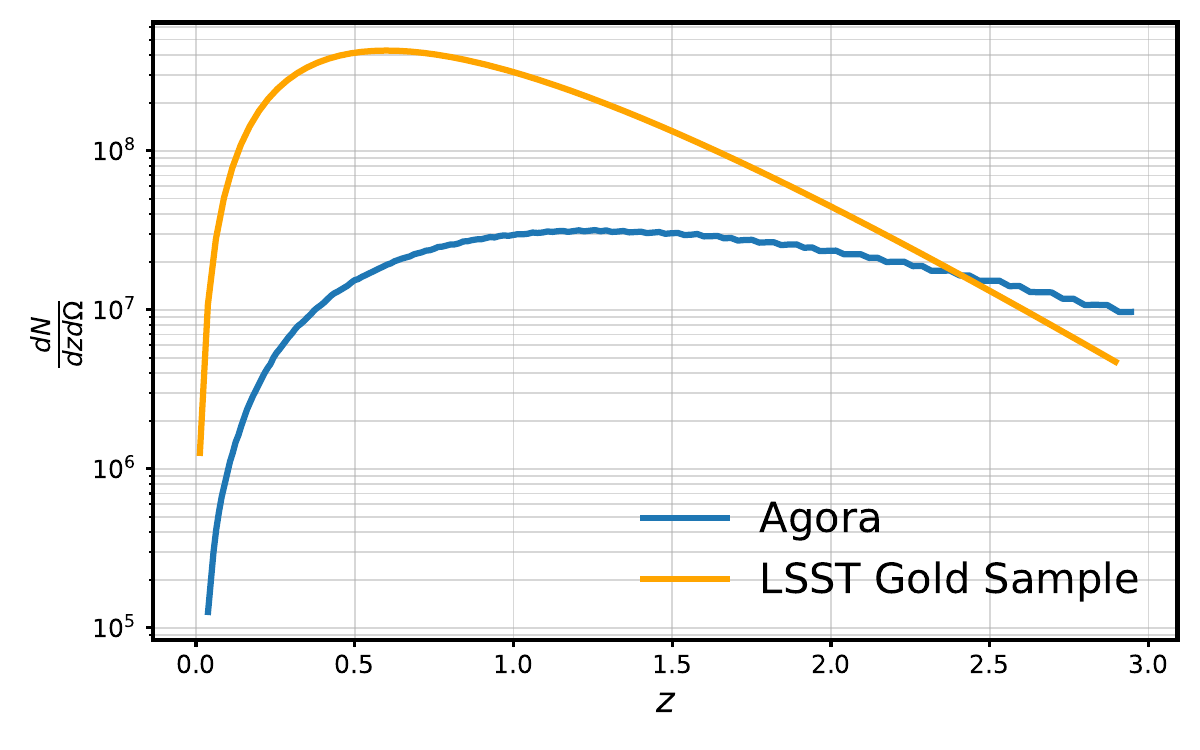}
\includegraphics[width=0.45\textwidth]{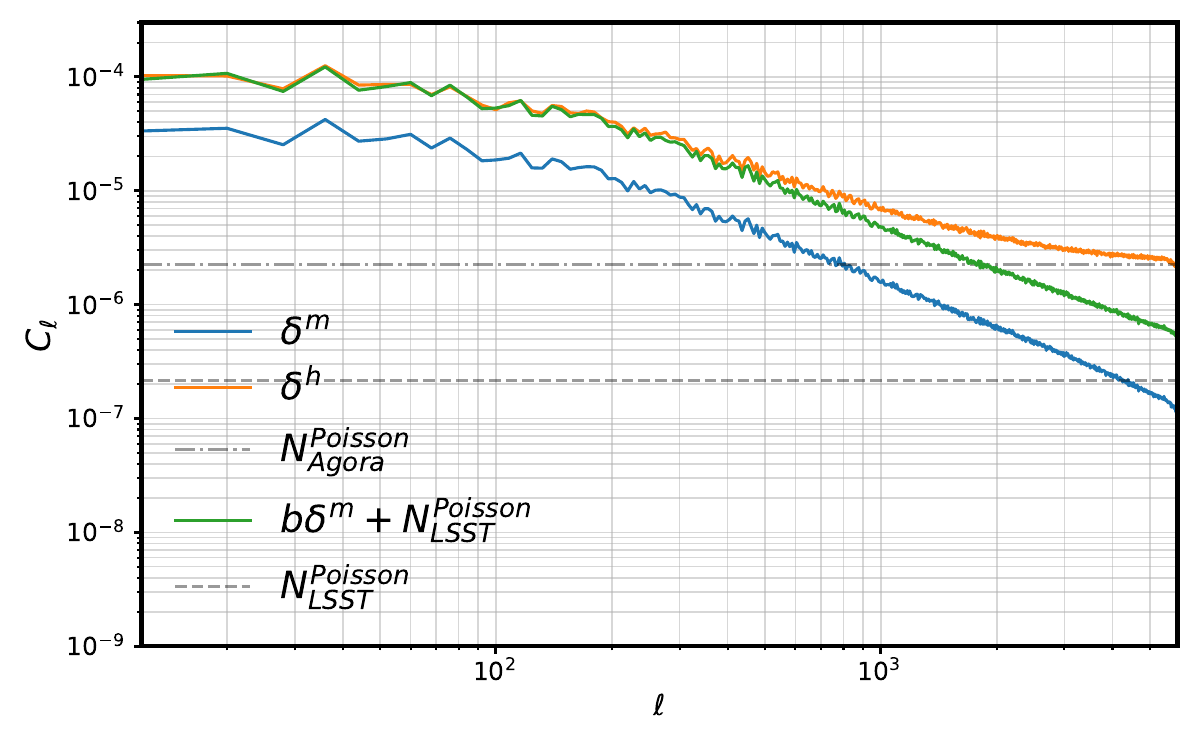}
\caption{Left: Halo density (Agora) and galaxy density (LSST Y-10) per sterad per redshift as a function of redshift. Right: Power spectra of halo overdensity from Agora and emulated halo overdensity (i.e. Poisson sampled biased matter overdensity) for LSST-Y10 gold sample galaxy density level. At the kSZ length scales of $\ell > 4000$ the Agora map is noise dominated while the emulated LSST map is signal dominated. }.\label{fig:ng}
\end{figure}

The second required simulation product is the halo/galaxy catalog. In Fig. \ref{fig:ng} (left) we show the differential number density of the 
    halo catalog from the Agora\footnote{https://yomori.github.io/agora/index.html} simulation, as well as the expected galaxy number density of the Rubin Observatories LSST-Y10 gold sample \cite{2009arXiv0912.0201L}. 
    We note that we used the lightcone halo catalog from Agora which is not tailored for a specific survey. Agora does provide a LSST-Y1 galaxy catalogue, but we are not using it in this analysis. 
Given the densities in Fig. \ref{fig:ng} (left), we thus expect weaker constraints with Agora halos (due to the mass threshold of the Agora halo catalogue of $10^{11}_\odot$ solar masses) than can be obtained with Rubin Observatory LSST-Y10. Unfortunately, as we shall find, to increase the signal-to-noise with the likelihood, a larger number density than Agora is required. For this reason, in addition to analyzing the actual halo catalog in Agora, we also test our analysis on Poisson sampled biased matter density maps for various shot noise levels. More precisely, we emulate higher-density halo maps by multiplying Agora matter densities by halo biases and adding Poisson shot noise $n^{-1}_g$ corresponding to the target density. While Poisson shot noise is certainly not a perfect approximation of small-scale halo formation, this allows us to approximately extend our analysis to very high galaxy densities. Fig. \ref{fig:ng} (right) shows the power spectrum of both the real Agora halo map as well as the emulated one with Rubin Observatory density. In all of our results below we do not include photo-z redshift errors, however, the redshift resolution is limited by the choice of 60 radial bin. 

The third required simulation product is the observed CMB map. While Agora comes with its own simulated CMB map, which includes properly correlated CMB lensing and various CMB secondary anisotropies, for our present purpose it is more convenient to generate our own CMB map. We generate primary CMB from the lensed CMB power spectrum of CAMB. We then add kSZ generated according to the binned kSZ model described in Sec. \ref{binned_ksz}, by multiplying the binned radial velocity maps with the binned electron density maps (assuming $\delta_e = \delta_m$) at full resolution and adding all 60 bins up to redshift 3. This is the kSZ generated by our simulation volume. The real data will include some additional kSZ. First, some kSZ will be generated by radial distances below our binning resolution (see \cite{Cayuso_2021}). Second, kSZ from higher redshifts including reionization kSZ will also contribute. To approximate both of these effects, we add \say{uncorrelated kSZ} to the CMB map, as described in Eq. \eqref{eq:totalcmb}. Below we both show results for the simulation kSZ alone, as well as for an added uncorrelated kSZ with the same power spectrum and amplitude as the simulation kSZ. We made this choice because high-z kSZ (here $z>3$ including reionization) may have a similar total amplitude as low-z kSZ \cite{Cai:2021hnb,Foreman:2022ves}.

\begin{figure}[t!]
\centering
\includegraphics[width=0.5\textwidth]{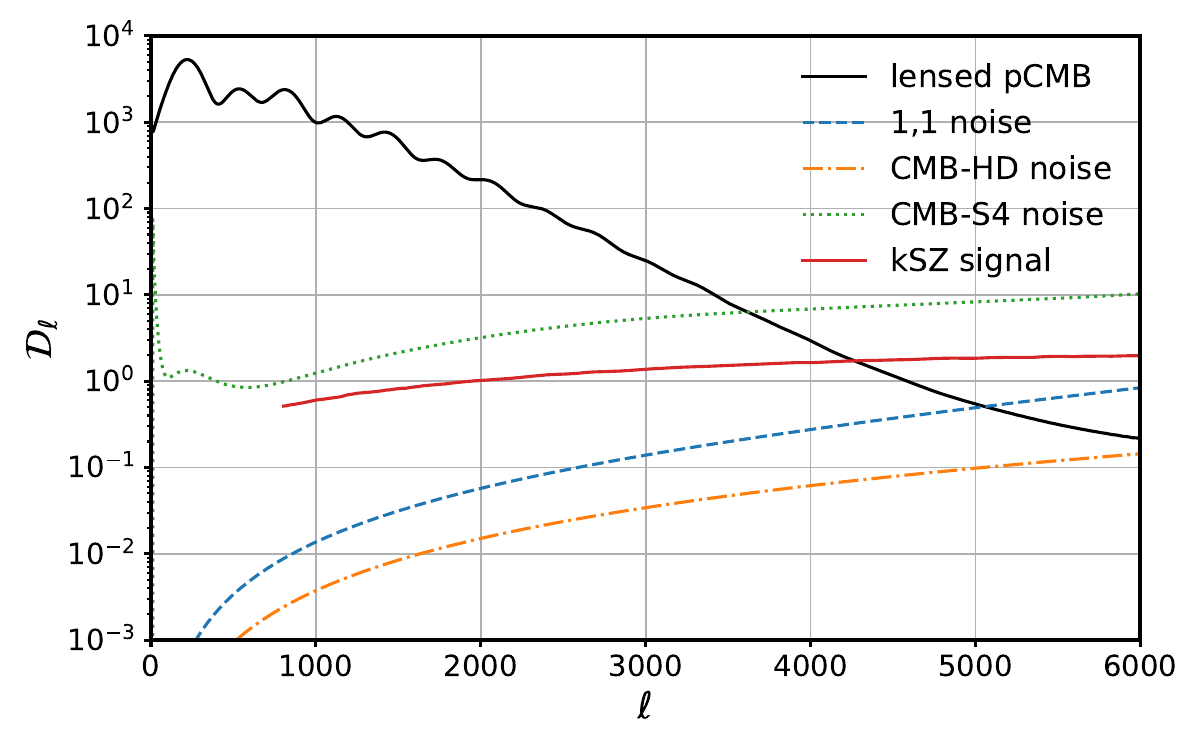}
\caption{Lensed CMB power spectrum along with three different instrumental noise levels and generated kSZ signal (simulation volume only). For CMB-S4 the kSZ (shown only for our simulation volume $0.5<z<3$) is below the noise level, while for the futuristic experiments, it is above the noise level. Here the CMB-S4 noise curve includes ILC foreground deprojection, while the $1,1$ noise means white noise with $N_T = 1 \mu K \mathrm{arcmin}$ and $\theta_{FWHM} = 1\ \mathrm{arcmin}$ (without ILC). For either of these experiments, our analysis uses only scales up to $\ell\simeq6000$ to limit the GPU requirements of this theoretical study.}\label{fig:cmbnoises}
\end{figure}

In our analysis, we consider three different CMB noise levels, as shown in figure \ref{fig:cmbnoises}. CMB-S4 noise corresponds to CMB-S4 ultra-deep ILC curve taken from \cite{ilcs4}. Finally, the curve named "$1,1$" was generated from the power spectrum $N(l) = N^2_Te^{\frac{l(l+1)\theta^2_{FWHM}}{8\ln 2}}$ with $N_T = 1\ \mu K \mathrm{arcmin}$ and $\theta_{FWHM} = 1\ \mathrm{arcmin}$  (without ILC foreground cleaning) to be comparable to the forecast in \cite{M_nchmeyer_2019}. The CMB-HD noise curve was generated according to \cite{Han_2022}
by co-adding instrumental noises from the frequency channels of 150 and 90 GHz. The level of foregrounds in 150 GHz and 90 GHz, that is left after foreground cleaning, is shown with a blue line in Figure 1 of  \cite{Han_2022}. We note that it is dominated by the reionization kSZ signal, the influence of which is discussed later in this work. The level of CIB foreground that is left is much less than the instrumental noise. Whether CMB-HD can reach such low foreground levels (using other frequency channels for foreground cleaning) is currently being investigated by the CMB-HD collaboration. In this work, we do not include correlated foreground residuals, which need to be studied in more detail even in the case of the standard quadratic estimator method. 
We show these noise spectra as well as the kSZ from our simulation volume in Fig. \ref{fig:cmbnoises}.

\subsection{Results for observed matter $\delta_m$ with Poisson shot noise}
\label{sec:resdeltam}

We now test our likelihood pipeline on the Agora simulations, assuming that we observe the Poisson noise corrupted biased matter field (see Fig. \ref{fig:ng} right) in addition to the CMB. We thus find the MAP to the posterior Eq. \eqref{eq:posteriorkszmatter}, which includes the likelihood Eq. \eqref{eq:Lkszmatter}. In the next section, we will then analyze the Agora halo catalog instead of the matter field. 

\begin{figure}[t!]
\centering
\includegraphics[width=1.\textwidth]{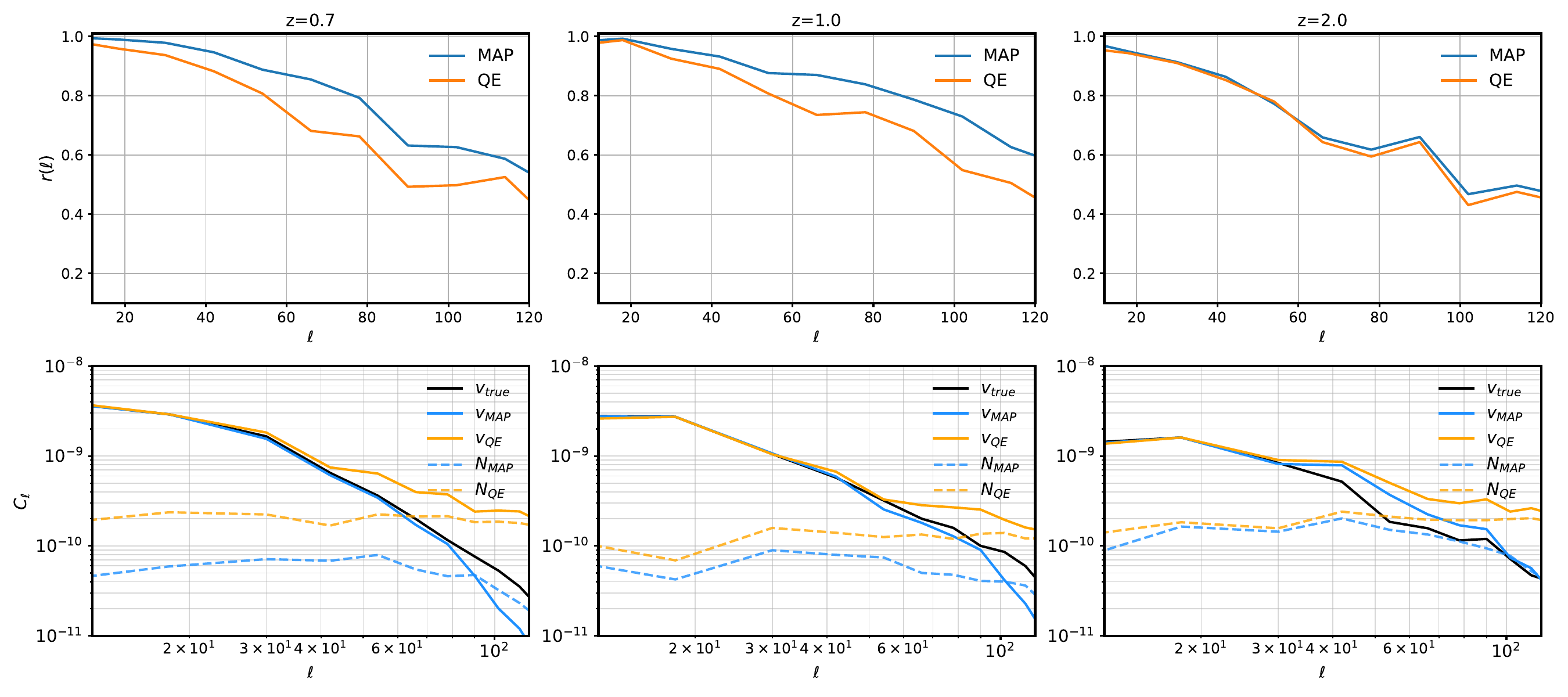}
\caption{Velocity reconstruction based on $\delta_m$ and $\theta^{CMB}$. Cross-correlation coefficient $r_{\ell}$ (upper row) and noise power spectra (lower row) of QE and MLE as a function of multipole number $\ell$ for three different redshifts. CMB noise and galaxy density are respectively: $N^{pCMB}_{1,1}$ ($N_T = 1 \mu K \mathrm{arcmin}$ and $\theta_{FWHM} = 1\ \mathrm{arcmin}$, without foregrounds) and $n^g_{LSST}$ (LSST gold sample). We include data in 60 redshift bins ($0.5<z<3$) with angular resolution $\ell_{max}=6000$ in the reconstruction, which is the resolution limit with our GPU memory.}\label{fig:rec_vel_matt_three_noises}
\end{figure}

We first consider the shot noise that corresponds to the $n_g$ prediction for Rubin Observatory (Fig. \ref{fig:ng} left), and the $N^{pCMB}=N^{pCMB}_{1,1}$ CMB noise as described in Sec. \ref{sec:experimentdata}. Note that we do not take into account photo-z errors (except by redshift binning), so the results for real Rubin Observatory galaxies would be somewhat weaker. Figure \ref{fig:rec_vel_matt_three_noises} shows the cross-correlation coefficient and reconstruction noise for quadratic estimator (QE) and maximum a posteriori (MAP) estimator for three different redshifts $z=0.7,\ 1,\ 2$. The reconstruction noise is defined as $N(\ell) = \langle|v^{true}-b\hat{v}|^2\rangle$ and the cross-correlation coefficient is $r(\ell) = \frac{\langle v^{true}\hat{v}\rangle}{\sqrt{\langle\hat{v}\hat{v}\rangle\langle v^{true}v^{true}\rangle}}$. For this configuration the MAP estimator performs significantly better than QE, especially at lower and intermediate redshifts, giving lower reconstruction noise and higher cross-correlation coefficient.

Let's discuss the curves in Fig. \ref{fig:rec_vel_matt_three_noises} in somewhat more detail. First, the quadratic estimator has lower noise at intermediate redshift $z=1$ than at both lower and higher redshift bins. This is expected because the volume of the $z=1$ bin is larger than at $z=0.7$, and the number density is larger at $z=1$ than at $z=2$ (see Fig. \ref{fig:ng}), resulting in the lowest noise at intermediate redshift. Second, cosmic variance fluctuations (due to having only a single simulation) between the QE and the MAP are similar, as expected. Third, the noise of the MAP (as well as the amplitude of the reconstructed velocity signal) goes down at large $\ell$, while the noise of the QE stays flat. This is the expected behavior because the MAP includes inverse variance weighting so that noisy scales are being suppressed by $S(S+N)^{-1}$ (where $S$ is the signal covariance and $N$ is the noise covariance). If we were to Wiener filter the QE result, we would get the same behavior, i.e. the QE reconstructed power spectrum would fall at large $\ell$. To estimate the unbiased power spectrum from the Wiener-filtered field, one would include a power spectrum transfer matrix, as is commonly done in power spectrum estimation. Finally, a somewhat surprising result is that the improvement of MAP over QE is larger at $z=0.7$ than at $z=1$, even though the QE performs better at $z=1$. However, it is difficult to predict the improvement factor analytically because it depends on several redshift-dependent factors, in particular the number density $n$ per sky angle (which first grows and then falls towards higher $z$) as well as the amplitude of the cross power spectrum $P_{g \tau}$ (which grows towards lower $z$), and perhaps the non-Gaussianity of the small-scale fields (which grows towards lower $z$). As a final note, we cannot exclude the fact that our likelihood is not fully converged on small scales, despite our best effort to achieve convergence (see Sec. \ref{sec:implementation}). In particular, in the high redshift bin $z=2$, at high $\ell$, the MAP power should be below the true velocity field power, due to the inverse variance weighting. However, on large scales, which are those most interesting for cosmology, convergence appears to be robust. 

In all plots of the reconstruction power spectrum, we have adjusted an overall (scale-independent) amplitude parameter to match the theory power. This is equivalent to fixing the \say{velocity bias} described in \cite{arxiv.1810.13423}, which also has to be marginalized over in the case of the QE. The velocity bias is physically associated with the unknown normalization of the electron density in the real universe. In simulations, it could be fixed by carefully tracking normalizations and the masking bias but we have not done this here. We also estimated the small-scale power spectrum that appears in the prior term from simulations. In the future, we aim to fit a parametrization of the small-scale power spectrum jointly with the field parameters to take into account baryonic uncertainty. 

We now explore the influence of different experimental configurations. As a futuristic case, we consider an experiment with 10 times the number density of Rubin Observatory. While no experiment with such a low shot noise is planned in the near term, this configuration allows us to probe how the signal-to-noise scales with galaxy number density. As described in Sec. \ref{sec:experimentdata} we also consider three different CMB noise configurations, CMB-S4 (ILC), \say{1,1 noise}, and CMB-HD. As before, we include data in 60 redshift bins up the $\ell_{max}=6000$ in the reconstruction, which is the resolution limit with our GPU memory (three GPUs with each 16GB). The absolute SNR is thus lower than could be achieved with a larger $\ell_{max}$ (especially for a CMB noise better than CMB-S4), but our main concern here is to demonstrate the improvement over the QE, for which our GPU setup suffices. For these configurations, we provide signal-to-noise improvement factors in Table \ref{tab:0}, ranging in values from 1 to 4. Unfortunately with CMB-S4 ILC noise curves we do not find an improvement. Note however that this result could be different with halos, a more complicated halo-based kSZ model, and halo mass weighting in the galaxy field, which we aim to explore in the future with a more high-resolution simulation. With the present simulation, the question of whether an improvement with CMB-S4 is possible cannot be definitively answered.  Comparing the $1,1$ noise case and the CMB-HD noise, we only find a substantial improvement in the $z=0.7$ bin, indicating that the galaxy shot noise is the limiting factor at higher redshifts for this extremely low noise configuration. In some bins, we find a slightly better result with the $1,1$ noise than with the HD noise, which may be due to small fluctuations in the convergence of our optimization procedure.

\begin{table}[tbh!]
    \centering
    \begin{tabular}{l|c|c|c}
    \hline
      & z=0.7 & z=1 & z=2 \\
      \hline
     $N_{1,1}^{pCMB}$ \& $n_g^{LSST}$  &3.2 &1.7 &1.3 \\
     $N_{1,1}^{pCMB}$ \& $n_g^{LSST}$ \& ukSZ   &2.2 &1.6 &1.1 \\
     \hline
     $N_{1,1}^{pCMB}$ \& $10\times n_g^{LSST}$ &3.4 &2.6 &1.6 \\
     $N_{1,1}^{pCMB}$ \& $10\times n_g^{LSST}$ \& ukSZ &2.8  &2.2 &1.5 \\
     \hline
     $N_{HD}^{pCMB}$ \& $10\times n_g^{LSST}$  &4.0 &2.7 &1.4 \\
     $N_{HD}^{pCMB}$ \& $10\times n_g^{LSST}$ \& ukSZ &2.9 &2.0 &1.4\\
    \end{tabular}
    \caption{Improvement of the likelihood over the QE for futuristic noise levels (reconstruction based on $\delta_m$ with Poisson shot noise). $N^{v}_{QE}/N^{v}_{MAP}$, averaged over $\ell=10-60$ at three different redshifts for various shot noise and CMB noises configurations. ukSZ means that we include an uncorrelated kSZ component of the same magnitude ($\mathcal{D}^{ukSZ}_{\ell=2000}=\mathcal{D}^{kSZ}_{\ell=2000} \approx 1 [\mu K]^2$) as the simulation kSZ to approximate kSZ contributions from outside the redshift range of the simulation. We include data in 60 redshift bins ($0.5<z<3$) with angular resolution $\ell_{max}=6000$ in the reconstruction. For CMB-S4 ILC noise we do not find an improvement.}
    \label{tab:0}
\end{table}

Next explore how additional kSZ power influences our results. In Fig. \ref{fig:rec_vel_matt_three_noises}, the CMB map contained the primary CMB, the simulation kSZ as well as the instrument noise. We now include an additional uncorrelated kSZ component as described in Sec. \ref{sec:experimentdata}, by adding it to the map and modifying the covariance of the CMB prior. This situation is more realistic, as our simulation generates kSZ only at the redshift of 0.5 to 3 but there is an additional kSZ signal coming from $z>3$ including the reionization epoch. Since the level of the reionization kSZ is not well known, we set it to be equal in power to the modeled kSZ signal ($\mathcal{D}^{ukSZ}_{\ell=2000}=\mathcal{D}^{kSZ}_{\ell=2000} \approx 1 [\mu K]^2$). We summarize the average improvement of MAP reconstruction with respect to QE, $N^{v}_{QE}/N^{v}_{MAP}$ both with and without adding unresolved kSZ signal in Table \ref{tab:0}. We find that the extra kSZ reduces the improvement factor, but significant improvement remains.

\begin{figure}[t!]
\centering
\includegraphics[width=0.45\textwidth]{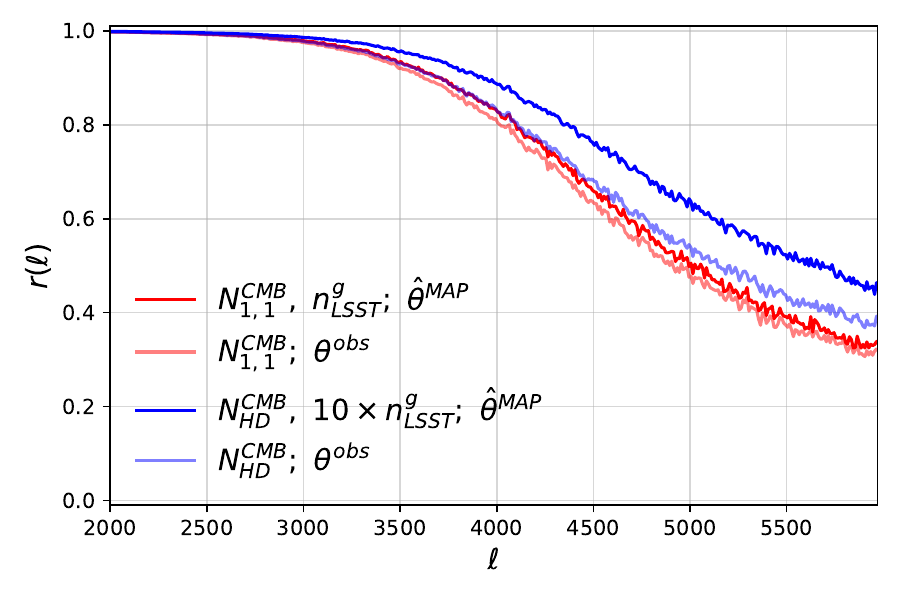}
\includegraphics[width=0.45\textwidth]{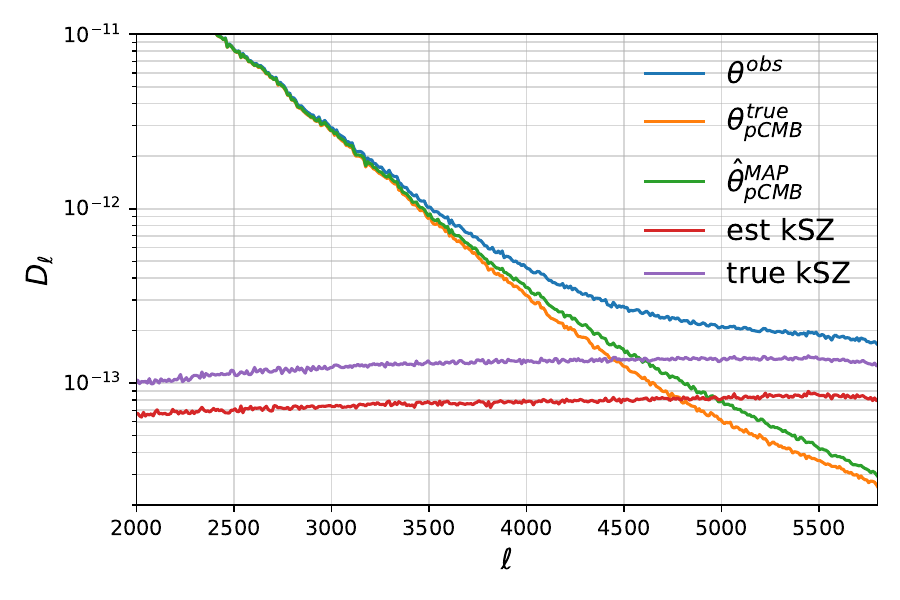}
\caption{Left: "De-kszing" of primary CMB. Cross-correlation coefficient $r(\ell)$ of $\hat{\theta}^{MAP}$ or $\theta^{obs}$ with $\theta^{pCMB}$ for two noise configurations: $N^{pCMB}_{HD},10\times n_g^{LSST}$ and  $N^{pCMB}_{1,1},n_g^{LSST}$. As expected, the reconstructed primary CMB cross-correlates better with the true primary CMB than the observed CMB does. The de-kSZing is not large here because we do not include an estimate of small-scale velocities from small-scale galaxies (see main text). Right: Power spectra of reconstructed and true kSZ signal (red and purple) and pCMB (green and orange) for the $N^{pCMB}_{HD},10\times n_g^{LSST}$ noise configuration (correspondent to blue curves on the left).}\label{fig:deksz}
\end{figure}

As an interesting consistency check, we now investigate whether our likelihood has succeeded in reconstructing the \say{kSZ-corrupted} primary CMB. The likelihood approach naturally provides a kSZ-reduced temperature anisotropy field. In figure \ref{fig:deksz} (left) we demonstrate this showing the cross-correlation coefficients $r(\ell)$ for $\theta^{pCMB}-\hat{\theta}^{MAP}$ and  $\theta^{pCMB}-\theta^{obs}$ for two MAPs obtained for two different noise configurations. We see that indeed the reconstructed primary CMB cross-correlates better with the true primary CMB than the observed CMB. In Fig. \ref{fig:deksz} (right) we show the reduced power in the reconstructed primary CMB map. Despite the futuristic experimental configuration, de-kSZing is not very efficient in our setup. This is because we only reconstruct velocities on large scales, while all velocity scales contribute to the total kSZ signal. For example in \cite{Foreman:2022ves} galaxies were used to make a template for velocities on small scales. It is possible to modify the likelihood to reconstruct the velocities jointly from the kSZ and the galaxy catalog, which would substantially improve de-kSZing. However, we leave this to future work as it involves non-linear modeling. 

\subsection{Results for observed halos $\delta_h$}
\label{eq:resultsdeltah}
In this section, we perform the same MAP analysis, but now using halos (rather than matter plus Poisson noise) from the AGORA simulations. The posterior is now given by Eq. \eqref{eq:Tgalposterior} and the likelihood is given by Eq. \eqref{eq:Tgallikeli}. A difference to the previous section is that we now observe the halo/galaxy field $\delta_g$, but we model the kSZ from $\delta_m = \delta_e$. As discussed in Sec. \ref{sec:dedgtemplate} we need to make a template for $\delta_e$ given observed $\delta_g$. In this section, we use the simplest possible template, given by Eq. \eqref{eq:electrontemplate}, with small-scale power spectra $P_{gg}$ and $P_{ge}$ estimated from the simulations.

\begin{figure}[t!]
\centering
\includegraphics[width=0.9\textwidth]{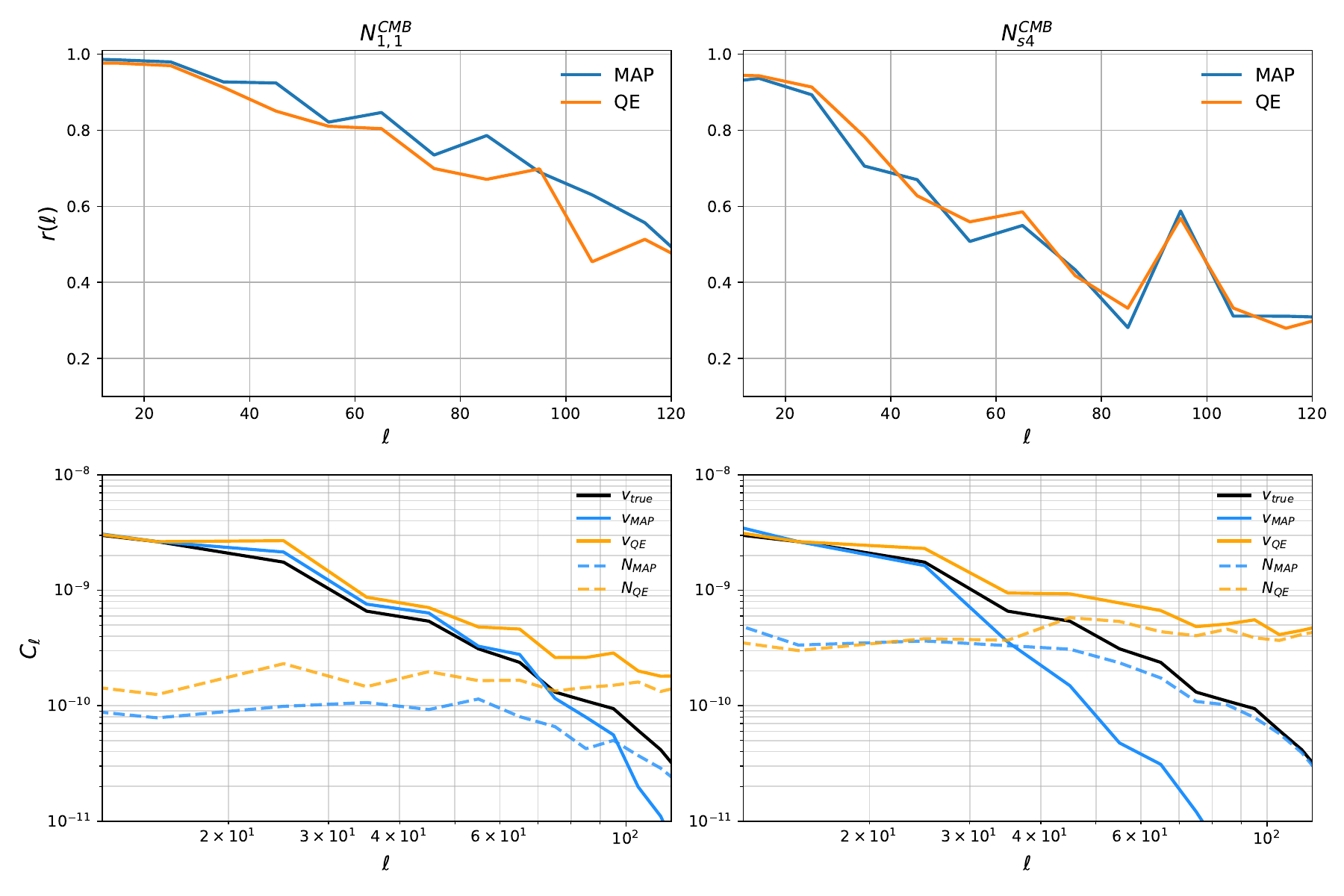}
\caption{Velocity reconstruction based on $\delta_h$ (Agora simulation halos) and $\theta^{CMB}$. Cross-correlation coefficient $r_{\ell}$ (upper row) and noise power spectra (lower row) of QE and MLE as a function of multipole number $\ell$ for two different CMB noise levels: $N^{pCMB}_{1,1}$(left) and 
$N^{pCMB}_{S4}$ (right) at $z=0.8$. In the reconstruction, we include data in 60 redshift bins ($0.5<z<3$) with angular resolution $\ell_{max}=6000$ in the reconstruction. For $N^{pCMB}_{1,1}$, we see an improvement by a factor of 2 while for $N^{pCMB}_{S4}$ (including ILC) we see no improvement.}\label{fig:5}
\end{figure}

Because the Agora halo density is lower compared to the LSST "gold sample" (Fig \ref{fig:ng}) by an order of magnitude, we expect less signal-to-noise and MAP improvement than in the previous section. The cross-correlation $r(\ell)$ and noise power spectra for reconstruction with halos and $N^{pCMB} = N^{pCMB}_{1,1}$ are shown on Figure \ref{fig:5} (left), finding about a factor of 2 improvement. Interestingly we find improvements even though the galaxy density is not as large as for LSST. The same plots but for higher CMB noise, $N^{pCMB} = N^{pCMB}_{S4}$ are depicted on Figure \ref{fig:5} (right). We see that at CMB-S4 noise, the MAP estimator performs equivalently to the QE in terms of signal-to-noise. 

\section{A machine learning-based estimator for the electron density}
\label{sec:machinelearning}

To estimate the velocity field, with both the QE or the likelihood, one needs an estimate or statistical connection between $\delta_e$ and $\delta_g$, as we discussed in Sec. \ref{jointlikeli}. In the previous section, we used the linear template estimate Eq. \eqref{eq:electrontemplate}, which is also the assumption usually made for QE reconstruction. However, in principle, this estimator can be an arbitrary non-linear function of $\delta_g$. In this section, we explore a neural-net-based estimator which we train on the Agora simulations. In a more complete exploration of this idea, one would train on hydro simulations to reconstruct a true baryon distribution, while on Agora we can only reconstruct $\delta_m \approx \delta_e$. We defer an exploration of this idea on CAMELS \cite{CAMELS:2020cof} or IllustrisTNG \cite{Nelson:2018uso} to future work.

Our goal is to train a neural network to increase the cross-correlation between the electron template derived from the galaxy field, and the true electron density field. Let us first show how this cross-correlation affects the reconstruction noise of the QE, where we have an analytic expression for the noise. The QE velocity reconstruction noise is sensitive to the cross-correlation between the true electron field and its tracer as follows:
\begin{equation}
(N^v_{\ell})^{-1} \propto \int dll \frac{(P_{\alpha}^{g\tau}(l))^2}{\tilde{P}^{\theta\theta}(l)\tilde{P}^{gg}(l)} = \int dll \frac{f^2_{\tau}r_{eg}^2(l)P^{ee}}{\tilde{P}^{\theta\theta}(l)},\ \ \ r_{eg}^2(l) = \frac{(P^{eg})^2}{\tilde{P}^{gg}P^{ee}}
\end{equation}
where we had $\delta_g$ as a tracer of $\delta_e$. Hence, we can reduce noise by having a tracer with better $r_{eg}^2$. In the region of high $\ell$, primary CMB is suppressed. If CMB noise is small, the cross-correlation coefficient of the true electron density field with its tracer defines noise completely. Indeed, assuming that at high $\ell$, $\tilde{P}^{\theta\theta}(l)\sim const$, we know that $P^{ee}(\ell) \sim \frac{\ell^2}{\chi^2_{*}}P^{ee}(|\bb{k}|=\frac{\ell}{\chi_{*}}) \sim const.\times \ell^{-1}$, so that $(N^v_{\ell})^{-1} \propto \int r^2(\ell)d\ell$. Hence, in the case of perfect reconstruction, the noise is only limited by resolution.

We train a 8-block ResNet \cite{he2016deep} with 3x3 convolutional layers on pairs of $(\delta_e,\delta_g)$ to minimize the standard RMS loss for the three cases of matter tracer used in the preceding sections. In our simulation $\delta_e = \delta_m$ and the matter tracer $\delta_g$ here is either Poisson noise corrupted $\delta_m$ with noise $N = 1/n_{g,LSST}$ and $N = 0.1/n_{g,LSST}$ or the Agora halo field. The network converges fast (the whole training takes a couple of minutes on RTX-A4000). We implement it with \texttt{jax} and \texttt{haiku} - a neural network library, compatible with \texttt{jax}. This allows us to effortlessly include this piece in likelihood optimization for velocity reconstruction (although the neural network template is equally applicable to the QE).

\begin{figure}[t!]
\centering
\includegraphics[width=0.9\textwidth]{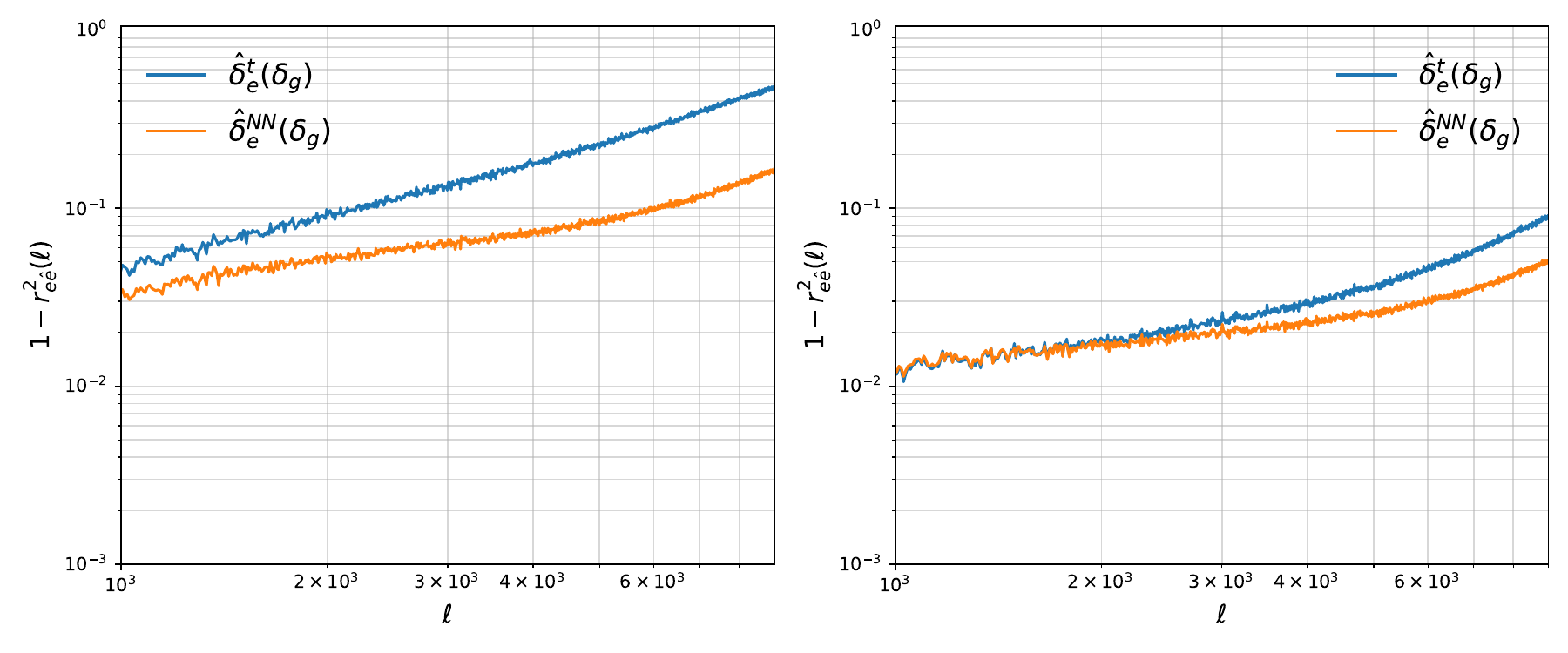}
\caption{Cross-correlation $1-r^2_{e\hat{e}}(\ell)$ of the true electron field with its estimate from the analytic template Eq. \eqref{eq:electrontemplate} compared to the neural network estimate for two halo density levels - $n_{g,LSST}$ (left) and $10\times n_{g,LSST}$ (right). We find a significant improvement in the cross-correlation due to the neural network, which turns out to be larger with higher shot noise (lower galaxy density). Note however that in our simulations $\delta_e=\delta_m$ and results may change with realistic hydrodynamic simulations and galaxy properties. }\label{fig:nn_el_reco}
\end{figure}
As can be seen in Figure \ref{fig:nn_el_reco}, NN estimator for electron density gives an improvement in cross-correlation, especially for higher multipoles, where the kSZ signal is observable. We don't see any improvement on the lower-density Agora halo catalogs. Instead of a ResNet, it would also be interesting to train a more physical low-parameter model such as Lagrangian Deep Learning \cite{Dai:2020ekz}. 

\begin{figure}[t!]
\centering
\includegraphics[width=0.9\textwidth]{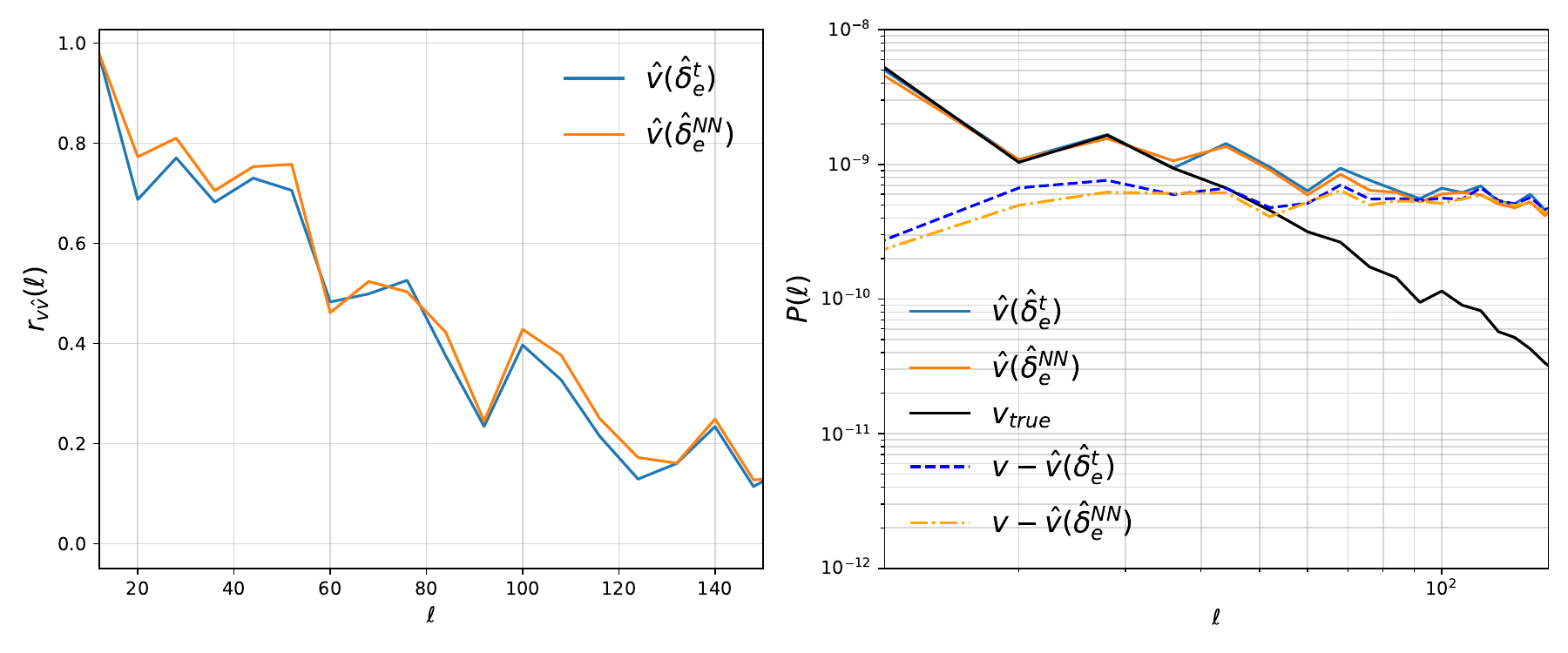}
\caption{Left: Cross-correlation coefficient $r_{v\hat{v}}(\ell)$ of true and QE-reconstructed velocity field based on the standard and the neural network estimate of the electron density. Right: Corresponding power spectra and residuals. The experimental parameters are similar to an LSSTxSO analysis. We find a small $25\%$ improvement when the QE uses the neural network template, rather than the analytic template (Eq. \eqref{eq:electrontemplate}). }\label{fig:qe_nn_temp}
\end{figure}

Next, we investigate how the improved electron template affects velocity reconstruction. Here we consider a more realistic case of high CMB noise - Gaussian white noise with $N_T=5\ \mu K \mathrm{arcmin}$ and $\theta_{FWHM}=1.5\ \mathrm{arcmin}$, which is closer to the target noise of Simons Observatory. We also take halo shot noise at the LSST level, so this forecast is oriented towards an LSSTxSO analysis. We use the QE velocity estimator, but at these noise levels, we expect the likelihood to perform likewise, as was seen in the previous section. Figure \ref{fig:qe_nn_temp} shows approximately a factor of $1.25$ improvement in reconstruction noise in low $\ell$ region resulting from our NN-based estimator of electron density applied to the QE. 

As we saw, in our setup the possible improvements due to the neural network are not very large. However, this conclusion could change with more realistic hydrodynamic simulations where the baryon density is being simulated, as well as by considering the properties (such as mass) of halos and galaxies. We plan to explore this question in more detail in future work on hydrodynamic simulations. One may also ask whether the neural network technique is sensitive to baryonic uncertainty. Here we point out that an analytic template, as used in the QE, is also not correct in nature. As shown in \cite{arxiv.1810.13423}, on large scales, such baryonic uncertainty can be included by marginalizing over a \say{velocity bias} on large scales. This is still true in the neural network case. All we require is that the neural network increases the cross-correlation coefficient in reality over the analytic template. Further, we did not take into account CMB foreground residuals, and a detailed study of our method should take them into account.

\section{Conclusion}
\label{sec:conclusion}

In this paper, we developed an auto-differentiable likelihood pipeline for the cross-correlation of the CMB and a large-scale structure survey due to the kSZ effect. As is the case for CMB lensing, which has been developed in more detail in the literature because of its larger signal-to-noise for current experiments, a likelihood pipeline is more statistically sensitive than the quadratic estimator at low experimental noises. Beyond the increased statistical sensitivity, a likelihood approach provides a general framework to fit a complete model to the data, which can include both physical and experimental parameters. 
Likelihood-based kSZ velocity reconstruction is computationally tractable because the relevant scales of the large-scale velocity field and the small-scale matter field can be separated to a good approximation, so non-linear differentiable forward modeling of structure formation is not necessary. We show that our approach is computationally tractable for a realistic survey size, and provide expected improvement factors over the QE for different idealized experimental configurations. We further developed a machine learning approach to the estimation of the electron density given observed galaxies, which may be able to improve the signal-to-noise for Simons Observatory, although a more detailed study using hydrodynamic simulations is required.  

There are two main directions of future work which we will explore. The first direction is to make our forecasts more realistic. Based on the results in the present work, we do not find significant gains in signal-to-noise at the resolution of CMB-S4 and Rubin Observatory (LSST). However, the available Agora simulations do not fully satisfy the required simulation constraints for this forecast, and to probe LSST density we made the simplified assumption that we observe the matter field of Agora with Poisson shot noise. In a simulation with realistic galaxies, with sufficient density, one could take into account galaxy properties, in particular tracers of the halo mass. Halo masses significantly reduce the shot noise in the galaxy field \cite{seljak2009suppress}, and could improve our forecast. Fortunately, more realistic high-resolution simulations are starting to become available. A more elaborate forecast would also take into account photo-z errors beyond the simple redshift binning we have performed here. 
A future study should also take into account foregrounds that correlate with large-scale structure and can affect both the QE results and the likelihood. In particular, the tSZ and CIB correlate with the galaxy field and their remnants (after foreground cleaning) could bias the reconstruction. Our improvement factors can be sensitive to these effects, however, this is a complicated topic that is beyond the scope of the present work. The second main goal for future work is the inclusion of cosmological and astrophysical parameters in the likelihood as outlined in Sec. \ref{sec:parameters}. Jointly fitting such parameters together with fields is in principle possible, but achieving convergence in the fit can be difficult. In addition, to obtain unbiased parameter measurements and error bars, one would need to integrate out the field level variables, perhaps with a Gaussian approximation around their MAP. Recent related work in the case of CMB lensing includes \cite{Millea:2021had,Millea:2020iuw,Bianchini:2022wte}. We will explore these questions in future work.

\section*{Acknowledgements}

We thank Matthew Johnson for comments on the manuscript and the members of the Simons Observatory SZ working group for comments on this work. MM acknowledges support from DOE grant DE-SC0022342. Support for this research was provided by the University of Wisconsin - Madison Office of the Vice Chancellor for Research and Graduate Education with funding from the Wisconsin Alumni Research Foundation. To obtain results quoted in this work, we extensively used \texttt{jax} \cite{jax2018github} for construction of the autodifferentiable posterior, as well as tools from DeepMind JAX Ecosystem \cite{deepmind2020jax}, such as \texttt{optax} for the optimization via gradient descent and \texttt{haiku} \cite{haiku2020github} for the construction of a NN-based estimator of electron density. We also benefited from \texttt{pixell} and Python implementation of HEALPix\footnote{http://healpix.sourceforge.net} \cite{Gorski:2004by} \texttt{healpy} for data preprocessing, along with \texttt{numpy}, \texttt{scipy}, \texttt{matplotlib} \cite{harris2020array, 2020SciPy-NMeth, Hunter:2007} for various calculations and plotting, and \texttt{camb} \cite{2011ascl.soft02026L} for theory computations.

\printbibliography


\appendix

\section{Relations between angular power spectrum $C_{\ell}$, flat-sky approximation $P(l)$ and power spectrum of the 3D field $P(k)$}
\label{appdx:a}
We review the known correspondence between multipole and flatsky coordinates, apply them to our case of binned flatsky coordinates, and calculate the radial velocity power spectrum.

\subsection{Equivalence of $C_{\ell}$ and $P(l)$}
\label{appdx:a1}
Given some  field on a 2D sphere $A(\bb{\hat{n}})$, we define its spherical harmonics transform coefficients as 
\begin{equation}
    a_{lm} = \int d\Omega Y^*_{lm}(\theta, \phi)A(\bb{\hat{n}})
\end{equation}
and if the field is rotationally symmetric, its angular power spectrum is defined, as usual
\begin{equation}
    \langle a_{l_1m_1} a^*_{l_2m_2}\rangle  = C_{l_1} \delta_{l_1l_2} \delta_{m_1m_2}
\end{equation}
If we consider a small enough solid angle, we can approximate that part of a sphere as flat. Without loss of generality, we can choose $\bb{\hat{z}}$ as radial direction. As in \cite{10.1111/j.1365-2966.2007.11747.x}, we may write then 
\begin{equation}
\hat{\bb{n}}_{\Omega} = \{\sin \theta \cos \phi, \sin \theta \sin \phi, \cos \theta\} \xrightarrow[\theta\xrightarrow{} 0]{} \bb{\hat{z}} + \bb{\alpha} = \{\theta \cos \phi,\theta \sin \phi,1\}
\end{equation}
So that
\begin{equation}
    a_{lm} \approx \int d^2\alpha Y^*_{lm}(\theta, \phi)A(\bb{\hat{n}})
\end{equation}
Now we introduce continuous flat-sky fields

\begin{align}
    a(\bb{l}) = \int e^{-i\bb{l}\bb{\alpha}}A(\bb{\alpha})d^2\alpha.
\end{align}
 with power spectrum $P(l)$ defined as 
 \begin{align}
 \langle a(\bb{l}_1)a^*(\bb{l}_2)\rangle = (2\pi)^2\delta^{(2)}(\bb{l}_1-\bb{l}_2)P(l_1)
 \end{align} 
We are going to show that defined this way, $P(l)$ is an exact continuous analog of $C_l$. Approximation from \cite{10.1111/j.1365-2966.2007.11747.x} tells us that
\begin{equation}
    Y_{lm} \xrightarrow[\theta\xrightarrow{} 0]{} \sqrt{\frac{l}{2\pi}}J(l\theta)e^{im\phi_{\alpha}}
\end{equation}
Using this formula and Jacobi-Anger expansion: $e^{i\bb{\alpha}\bb{l}} = e^{il\theta\cos(\phi_\alpha-\phi_{l})}=\sum_n(i)^nJ_n(l\theta)e^{in(\phi_\alpha-\phi_l)}$ we can write:
\begin{align}
    a(\bb{l}) = \sum_n (-i)^n \int A(\bb{\alpha})d^2\alpha J_n(l\theta)e^{-in(\phi_\alpha-\phi_l)}
    \hspace{1cm} a_{lm} = \int A(\bb{\alpha})d^2\alpha J_n(l\theta) \sqrt{\frac{l}{2\pi}}e^{-im\phi_\alpha}
\end{align}
Comparing two formulas and employing an orthogonality relation $\int\frac{d\phi}{2\pi}e^{i\phi(m-n)} = \delta_{mn}$, we can finally get:
\begin{align}
a(\bb{l}) = \sqrt{\frac{2\pi}{l}}\sum_m (-i)^ma_{lm}e^{im\phi_l} \hspace{1cm} a_{lm} = \sqrt{\frac{l}{2\pi}}i^m\int\frac{d\phi_l}{2\pi}a(\bb{l})e^{-im\phi_l}
\end{align}
 Then 
 \begin{align}
     \langle a_{l_1m_1} a^*_{l_2m_2}\rangle &= \sqrt{\frac{l_1}{2\pi}}\sqrt{\frac{l_2}{2\pi}}\int\frac{d\phi_{l_1}}{2\pi}\int\frac{d\phi_{l_2}}{2\pi} \langle a(\bb{l}_1)a^*(\bb{l}_2)\rangle e^{im_1\phi_{l_1}}e^{-im_2\phi_{l_2}} \\
     & = \sqrt{\frac{l_1}{2\pi}}\sqrt{\frac{l_2}{2\pi}}\int d\phi_{l_1}P(l_1)e^{i\phi_{l_1}(m_1-m_2)}\frac{\delta(l_1-l_2)}{l_1}\\
     &=P(l_1)\delta_{l_1,l_2}\delta_{m_1m_2}
 \end{align}
 The other way round, the result may be obtained if we formally consider high $l$, so that $\sum^{l}_{m=-l} e^{im\phi_l} \approx \sum^{\infty}_{m=-\infty}e^{im\phi_l}=2\pi\delta(\phi_l)$:
 \begin{align}
     \langle a(\bb{l}_1)a^*(\bb{l}_2)\rangle & = \sum_{m_1}\sum_{m_2}(-i)^m_1i^m_2\sqrt{\frac{2\pi}{l_1}}\sqrt{\frac{2\pi}{l_2}}\langle a_{l_1m_1} a^*_{l_2m_2}\rangle \\
     &=\sum_{m_1}\frac{2\pi}{l_1}\delta_{l_1,l_2}e^{im_1(\phi_{l_1}-\phi_{l_2})}C_{l_1} \\
     &\approx (2\pi)^2C_{l_1}\frac{\delta(l_1-l_2)}{l_1}\delta(\phi_{l_1}-\phi_{l_2}) = (2\pi)^2C_{l_1}\delta(\bb{l_1}-\bb{l_2})
 \end{align}
 In these derivations, we made associations of discrete and continuous delta-functions, mainly $\delta_{l_1,l_2} = \delta(l_1-l_2)$ and considered the expression of delta-function in polar coordinates: $\delta(\bb{l})=\frac{\delta(l)}{l}\delta(\phi_l)$ 
\subsection{Relation between $C_{\ell}$ and $P(k)$}
\label{appdx:a2}
We define redshift-binned fields via integrals with a window function:
\begin{equation}
    f^i_{lm} = \int dzW_{z^i}(z)\int d\Omega_xY^*_{lm}(\hat{\bb{x}})f(\bb{x},z)
\end{equation}
In our case, the window function is a top-hat filter: $W_{z^i}(z) = \frac{1}{\Delta_z}(\theta(z+(z^i+\frac{\Delta_z}{2}){})-\theta(z-(z^i-\frac{\Delta_z}{2})))$ Here $\theta(z)$ is a Heaviside step function. Then for the density-like field $\delta^{m}$, it's easy to get
\begin{align}
    C^{\delta\delta}_{l}(z^i,z^j)= \frac{2}{\pi}\int dzdz' W_{z^i}(z)W_{z^j}(z')\int dk k^2P_{mm}(k)j_l(k\chi(z))j_l(k\chi(z'))
\end{align}
where we have used the Rayleigh expansion: $e^{i\bb{k}\bb{x}} = 4\pi\sum_{lm}i^lj_l(k\chi)Y_{lm}(\hat{\bb{k}})Y^*_{lm}(\hat{\bb{x}})$. For velocity $\bb{v}(\bb{x})$, we have a continuity equation that connects it to density mode on linear scales $\bb{v}(\bb{k})=\frac{ifaH}{k}\delta^m(\bb{k})\hat{\bb{k}}$. To compute radial velocity angular power spectrum, we define 
\begin{align}    
   v^r_{lm}(z^*)&=-i\int dzW_{z^*}(z)\int d\Omega_xY^*_{lm}(\hat{\bb{x}})(\bb{v}(\bb{x})\hat{\bb{x}}) \\
   &= \int dzW_{z^*}(z)\int d\Omega_xY^*_{lm}(\hat{\bb{x}})\int \frac{d^3k}{(2\pi)^3}\frac{faH}{k}\delta^m(\bb{k})(\hat{\bb{k}}\hat{\bb{x}})e^{i\bb{k}\bb{x}}
\end{align}
We can notice that $(\hat{\bb{k}}\hat{\bb{x}})e^{i\bb{k}\bb{x}} = \frac{\partial}{\partial(k\chi)}e^{i\bb{k}\bb{x}} = \frac{\partial}{\partial(k\chi)}e^{ik\chi\hat{\bb{k}}\hat{\bb{x}}}$ So that radial velocity angular correllation function is
\begin{equation}
    C^{vv}_{l}(z^i,z^j) = \frac{2}{\pi}\int dzdz' W_{z^i}(z)W_{z^j}(z')\int dk f(z)a(z)H(z)(f(z')a(z')H(z'))P_{mm}(k)j^{'}_l(k\chi(z))j^{'}_l(k\chi(z'))
\end{equation}
Here $j_l(x)$ and $j^{'}_l(x)$ are spherical Bessel function of the first kind and its derivative correspondingly.

\section{Error bars from the Hessian with the velocity-only kSZ likelihood}
\label{app:errorbars}

As we discussed in Sec. \ref{subsubsec:errors}, auto-differentiation makes it easy to evaluate the Hessian which can be used to obtain errorbars under the approximation that the posterior is Gaussian around the MAP. For our full posterior, including the small-scale fields, we cannot evaluate the full Hessian. We will explore methods to integrate out the small-scale fields in future work. Here instead we consider the simpler likelihood of Sec. \ref{sec:Lvonly} Eq. \eqref{eq:Pkszvel} where the only degrees of freedom are the velocity modes $v(\bb{k})$. We further simplify the problem to a single redshift bin, and make all fields Gaussian and periodic (no mask). We consider the case where all kSZ signal comes from one bin with known electron density so that the optical depth field $\tau$ is fully known. While in this case in principle there is an analytic MAP estimator, as discussed in Sec. \ref{sec:analyticMAP}, we instead use optimization to find the MAP as in our main analysis.

To obtain error bars after finding the MAP, we evaluate the Hessian. For a complex random field with symmetry $v(\bb{k}) = v(-\bb{k})^\dag$, we have:
\begin{equation}
\label{eq:complexfisher}
        cov(v(\bb{k})v(\bb{k'})) = \frac{1}{2} \left( \frac{-\partial^2 \ln \mathcal{P}}{\partial v(\bb{k}) \partial v(\bb{k'})^\dag} \right)^{-1} \Bigg|_{v=\hat{v}}
\end{equation}

\begin{figure}[t!]
\centering
\includegraphics[width=0.5\textwidth]{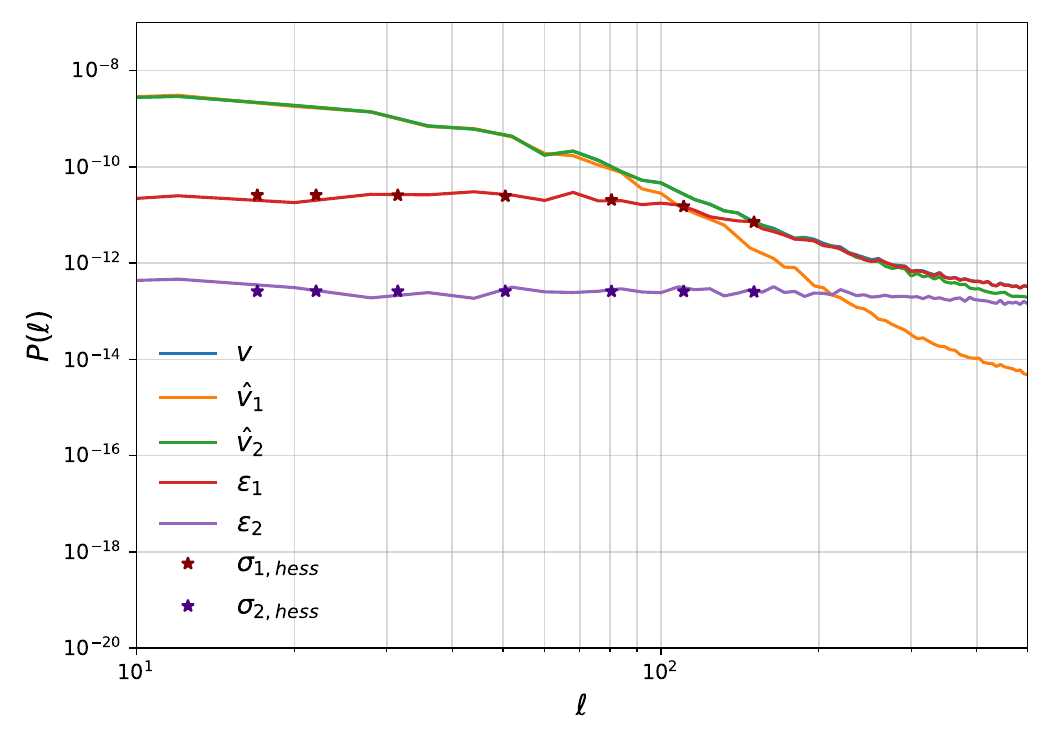}
\caption{Power spectra of MAP-estimated velocities $\hat{v}_{1,2}$ and residuals with the true fields $\epsilon_{1,2}$ for two levels of generated kSZ signal along with estimated errors from the Hessian. The two errors match very well, i.e. the Hessian errors are identical to the true residual errors.}\label{fig:hess_errors}
\end{figure}

We show the results of the MAP reconstruction in Fig. \ref{fig:hess_errors}. It depicts power spectra of MAP estimator $\hat{v}$ for two different levels of generated kSZ signal, power spectra of residuals $\epsilon = v_{true}-\hat{v}$, and analytic errors $\sigma_{hess}$. Analytic errors were evaluated via inversion of Hessian at MAP according to Eq. \eqref{eq:complexfisher}. Because reconstructed modes are (nearly) independent, Hessians can be computed in narrow bins of $\ell$, which significantly simplifies the evaluation and inversion process. Depicted values correspond to an average over the $\ell$ bins. We see that the error bar from the Hessian matches the true residual error very well. 
\end{document}